\documentclass[aps,prd,groupedaddress,preprint,tightenlines,eqsecnum,nofootinbib,showpacs]{revtex4}

\usepackage{graphicx,epsf,amssymb,amsbsy,amsfonts,amssymb,amsmath}
\usepackage[dvips]{color}

\newcommand{\eq}{\begin{equation}}
\newcommand{\eqe}{\end{equation}}

\newcommand{\eqa}{\begin{eqnarray}}
\newcommand{\eqae}{\end{eqnarray}}

\newcommand{\e}{\epsilon}
\newcommand{\eps}{\epsilon}

\def\vph{\vphantom{\db}}

\def\sect#1{section~{\ref{#1}}}

\def\eqn#1{eq.~(\ref{#1})}
\def\eqns#1#2{eqs.~(\ref{#1}) and (\ref{#2})}
\def\fig#1{fig.~{\ref{#1}}}

\def\app#1{appendix~{\ref{#1}}}

\def\Neqfour{{{\cal N}=4}}
\def\NeqFour{{{\cal N}=4}}

\def\NeqEight{{{\cal N}=8}}
\def\NeqOne{{{\cal N}=1}}

\def\nn{\nonumber}
\def\oneloop{{1 \mbox{-} \rm loop}}
\def\twoloop{{2 \mbox{-} \rm loop}}
\def\tree{{\rm tree}}

\def\spa#1.#2{\left\langle#1\,#2\right\rangle}
\def\spb#1.#2{\left[#1\,#2\right]}
\def\spash#1.#2{\spa{\smash{#1}}.{\smash{#2}}}
\def\spbsh#1.#2{\spb{\smash{#1}}.{\smash{#2}}}
\def\sand#1.#2.#3{%
\left\langle\smash{#1}{\vphantom1}^{-}\right|{#2}%
\left|\smash{#3}{\vphantom1}^{-}\right\rangle}
\def\sandpp#1.#2.#3{%
\left\langle\smash{#1}{\vphantom1}^{+}\right|{#2}%
\left|\smash{#3}{\vphantom1}^{+}\right\rangle}
\def\sandpm#1.#2.#3{%
\left\langle\smash{#1}{\vphantom1}^{+}\right|{#2}%
\left|\smash{#3}{\vphantom1}^{-}\right\rangle}
\def\sandmp#1.#2.#3{%
\left\langle\smash{#1}{\vphantom1}^{-}\right|{#2}%
\left|\smash{#3}{\vphantom1}^{+}\right\rangle}

\def\pol{\epsilon}
\def\da{{\vphantom{\dot{b}}\dot a}}
\def\db{{\dot b}}
\def\dc{{\vphantom{\dot{b}}\dot c}}
\def\dd{{\vphantom{\dot{b}}\dot d}}
\def\de{{\vphantom{\dot{b}}\dot e}}

\def\tsigma{\tilde \sigma}
\def\tlambda{\tilde \lambda}

\def\lfour{{}^{(4)}l}
\def\bu{\mathbf{u}}
\def\bw{\mathbf{w}}
\def\btu{\mathbf{\tilde{u}}}
\def\btw{\mathbf{\tilde{w}}}
\def\pfour{{}^{(4)}p}
\def\tm{\tilde{m}}

\def\shift{r}

\def\E{{\cal E}}
\def\nolsum{\sum\nolimits}

\newbox\charbox
\newbox\slabox
\def\s#1{{      
        \setbox\charbox=\hbox{$#1$}
        \setbox\slabox=\hbox{$/$}
        \dimen\charbox=\ht\slabox
        \advance\dimen\charbox by -\dp\slabox
        \advance\dimen\charbox by -\ht\charbox
        \advance\dimen\charbox by \dp\charbox
        \divide\dimen\charbox by 2
        \raise-\dimen\charbox\hbox to \wd\charbox{\hss/\hss}
        \llap{$#1$} }}

\begin{document}

\hbox{
SU-ITP-10/29 \hskip 10 cm  UCLA/10/TEP/107
}

\title{Generalized Unitarity and Six-Dimensional Helicity}

\author{Zvi Bern$^a$, John Joseph Carrasco$^b$, Tristan Dennen$^a$, 
 Yu-tin Huang$^a$ and Harald Ita$^a$ \\ $\null$}

\affiliation{${}^a$ Department of Physics and
  Astronomy\\ UCLA, Los Angeles, CA 90095-1547, USA\\
$\null$ \\
${}^b$  Department of Physics, Stanford University\\
  Stanford, CA 94305-4060, USA\\
$\null$\\
}

\begin{abstract}
We combine the unitarity method with the 
six-dimensional helicity formalism of Cheung and O'Connell 
to construct loop-level scattering amplitudes. 
  As a first example, we 
construct dimensionally regularized 
QCD one-loop four-point amplitudes.  As a nontrivial multiloop 
example, we confirm that the recently constructed four-loop 
four-point amplitude of $\NeqFour$ super-Yang-Mills theory, 
including nonplanar contributions, is valid for dimensions $D\le 6$.  
We comment on the connection of our approach to the recently 
discussed Higgs infrared
regulator and on dual conformal properties in six
dimensions. 
\end{abstract}

\pacs{12.38.-t, 12.38.Bx, 13.87.-a, 14.70.-e \hspace{1cm}}
\maketitle


\section{Introduction}
\label{IntroductionSection}

Recent years have seen a renaissance in the study of scattering
amplitudes.  This has been driven in part by the need for reliable
next-to-leading-order QCD calculations for
experiments at the Large Hadron Collider. It has also been driven by
the realization of profound hidden structures in scattering amplitudes
whose full import remains to be unraveled.  A key idea for many of the
advances has been that new amplitudes should be constructed using only
simpler on-shell
amplitudes~\cite{UnitarityMethod,BCFW}, avoiding
unphysical gauge-dependent quantities in intermediate steps.  The
on-shell viewpoint has revealed a number of remarkable structures in
gauge theories, including a description of amplitudes in terms of
curves in twistor space~\cite{WittenTopologicalString}, and a duality
between color and kinematics~\cite{BCJ}.  Maximally supersymmetric
planar amplitudes have an even richer structure, for example all-order
resummations of planar amplitudes valid at both weak and strong
coupling~\cite{BDS, AldayMaldacena}, new
symmetries~\cite{DualConformal,FourLoopCusp,DualConfWI} and a
Grassmannian structure~\cite{Grassmannian}.  On the practical side,
on-shell methods have allowed ever more complex calculations in
next-to-leading-order QCD~\cite{BlackHatEtc}, supersymmetric gauge
theories~\cite{FiveLoop,N4MultiloopExamples,TwoLoopSixPoints,
NimaBCFW} and supergravity theories to be carried out~\cite{GravityThree,
CompactThree, GravityFour, GravityReviews}.

The unitarity method~\cite{UnitarityMethod, ZFourPartons, BernMorgan,
  DDimUnitarity, BCFUnitarity, Buchbinder, BrittoDDim, OPP,
  Forde,Kilgore, FiveLoop, CachazoSkinner, GKM,Badger} gives a means
for constructing complete loop amplitudes directly from on-shell
tree amplitudes.  These methods are efficient and display a
relatively tame growth in complexity with increasing number of
external particles or loops, especially when compared to traditional
Feynman-diagrammatic approaches. With this approach, it is usually
advantageous to simplify the input tree amplitudes as much as
possible prior to applying them in loop calculations.  A key tool
for simplifying massless tree amplitudes strictly in four dimensions
has been spinor helicity~\cite{SpinorHelicity,TreeReview}.  However,
if we use dimensional regularization or any form of massive
regularization, massless four-dimensional spinor helicity techniques
are no longer directly applicable within the loops.

In this paper we will describe a unitarity-based approach using the
six-dimensional spinor-helicity formalism of Cheung and
O'Connell~\cite{CheungOConnell} to avoid these limitations.  As in
four dimensions, the six-dimensional spinor-helicity formalism offers
a relatively compact representation of amplitudes, making manifest
little-group properties. Here we are concerned with two basic
applications: one-loop QCD amplitudes and multiloop scattering
amplitudes in $\NeqFour$ super-Yang-Mills (sYM) theory.  Although
related, there are differences in the technical demands of the two
cases.  In particular, the QCD amplitudes have both ultraviolet and
infrared divergences, while the $\NeqFour$ amplitudes have only the
latter. All these divergences need to be regularized.  On the other
hand, the supersymmetric theory needs an efficient formalism to
organize its spectrum of states, especially at high-loop orders, in
much the same way as done in four dimensions using an on-shell
superspace~\cite{FourDOnShellSuperSpace,DeltaIdentity,FourDSupersums,SuperSum}.
As we discuss, both aspects can be addressed using six-dimensional
spinor helicity.

In the past, the main bottleneck for carrying out
next-to-leading-order QCD calculations.  has been the difficulty of
evaluating one-loop amplitudes~\cite{LesHouches}.  The unitarity
method offers a universal solution to this difficulty that scales
well with the number of external legs.  Any contributions which can be
captured by four-dimensional techniques --- the so called
``cut-constructible pieces'' --- may be computed efficiently using
four-dimensional helicity states in the cuts~\cite{UnitarityMethod,
BCFUnitarity, Forde, BlackHatI}.  The remaining rational contributions
tend to be the most complex (and time consuming) parts of
calculations, though there are a number of techniques for
dealing with such pieces.  In the bootstrap approach, the rational
pieces are constructed by on-shell recursion in four
dimensions~\cite{OneLoopRecursion,BlackHatI}.  Another approach uses
$D$-dimensional generalized
unitarity~\cite{BernMorgan,DDimUnitarity,TwoLoopSplit,BrittoDDim,Kilgore,GKM}.
A six-dimensional unitarity procedure for any one-loop QCD amplitude
has been given in ref.~\cite{GKM} using the on-shell procedure of
reducing integrals in ref.~\cite{OPP}.  A related approach makes use
of the relation between massive states in four dimensions and those in
extra dimensions to reduce the integrals obtained from unitarity
cuts~\cite{BernMorgan,Badger}.  The six-dimensional helicity approach
described here is well suited for either of the latter two approaches
for carrying out integral reductions.  A convenient means for doing so
is by reexpressing the spinors of our six-dimensional approach as
four-dimensional spinors and what are effectively mass parameters.
This gives expressions similar to those used in Badger's
four-dimensional massive approach~\cite{Badger}.  We illustrate this
using some simple one-loop four-gluon contributions to QCD
amplitudes~\cite{BernMorgan}.

We will also consider multiloop $\NeqFour$ sYM amplitudes.  To set
this up, we make use of the six-dimensional on-shell superspace
constructed by two of the authors and Siegel (DHS)~\cite{DHS}.  General constructions of on-shell superspaces have
  been discussed in ref.~\cite{Boels}. In
strictly four dimensions we have a well-developed on-shell
superspace~\cite{FourDOnShellSuperSpace,DeltaIdentity,FourDSupersums,SuperSum}
for tracking contributions from different states in the multiplet.
However,as already noted above this leaves open the question of
whether contributions are missed by using four-dimensional momenta in
the unitarity cuts or in the recent BCFW constructions of planar loop
integrands~~\cite{NimaBCFW,BoelsBCFW}.  This is especially important
when constructing amplitudes for use in $D>4$ studies, but even in
dimensionally regularized four-dimensional expressions, when
divergences are present, there can be nonvanishing contributions from
terms that vanish naively in four dimensions.  This becomes more
important as the loop order or the number of external legs grows,
allowing for a greater number of potentially problematic terms.
Besides terms with explicit dependence on extra-dimensional momenta,
such terms can be formed from antisymmetric combinations momenta which
vanish in four dimensions.  For example, we know that
dimensionally regularized two-loop six-point amplitudes in $\NeqFour$
sYM theory have such terms~\cite{TwoLoopSixPoints}. In theories with
fewer supersymmetries, such terms occur with more frequency, and, for
example, appear in one-loop QCD amplitudes~\cite{BernMorgan}.

To illustrate the supersymmetric formalism, we first describe the
three-particle cut of the planar two-loop four-point $\NeqFour$
amplitude~\cite{BRY}, before turning to the rather nontrivial case of
four-loop four-point amplitudes in this theory, including the
nonplanar terms.  The latter amplitude has recently been
computed~\cite{FourLoop}, using mainly four-dimensional techniques.
Direct calculations of four-point gluon amplitudes in $\NeqOne$ sYM in
$D=10$ dimensions, which upon dimensional reduction gives $\NeqFour$
sYM theory, confirm that all terms are captured by calculating
directly in four dimensions, through three loops.\footnote{This
property is special to maximal supersymmetry, and we have no
expectation that it should hold for other theories.}  At four loops,
similar $D=10$ checks have been performed in the planar case, up to a
mild assumption that no term has a worse power count than the
amplitude~\cite{FourLoopCusp}.  However, in the case of nonplanar
contributions, there is no complete check that the expressions built
using four-dimensional momenta in the cuts are complete, although a
number of strong consistency checks have been performed~\cite{FourLoop}. In
this paper we confirm that the expressions for the amplitudes of
ref.~\cite{FourLoop} are valid for $D\le 6$, as expected.

The higher-dimensional viewpoint offers a simple way to
introduce a gauge-invariant massive regulator. If we do not integrate
the loop momenta over the extra-dimensional momenta, but integrate
over only a four-dimensional subspace, we obtain a massive infrared
regulator, where the masses are effectively the extra-dimensional
momenta. This allows us to connect to the massive Higgs regulator
recently developed by Alday, Henn, Plefka and
Schuster~\cite{HiggsReg}, which offers a number of advantages over
dimensional regularization for planar $\NeqFour$ sYM amplitudes.
Although it seems likely that terms proportional to the mass are
suppressed as $m\rightarrow 0$~\cite{NimaBCFW}, such mass terms are important for
studies of the Coulomb branch of $\NeqFour$ sYM theory. In many cases
it is straightforward to convert the previously determined
dimensionally regularized massless integrands to massive
forms~\cite{HiggsReg}. Nevertheless, one would like a means for
obtaining or confirming such integrands directly via unitarity to
ensure their reliability or for deriving regulators valid for the
nonplanar case.  The six-dimensional on-shell superspace approach
gives us a convenient means for doing this.  

At four points, the integrands of $\NeqFour$ sYM
integrands do not depend explicitly on the space-time dimension, but
only implicitly through Lorentz dot products, as already noted in
refs.~\cite{BRY, BDDPR, GravityThree,CompactThree, FourLoopCusp,
GravityFour, FourLoop} and explicitly confirmed here at four loops.
This suggests that the dual conformal properties, which impose strong
constraints on the form of the integrands in four dimensions, will
impose similar constraints in higher dimensions.  In addition,
dual conformal invariance can be extended to the massive
Higgs-regulated case~\cite{HiggsReg}.  These facts suggest that the
dual conformal properties of planar $\NeqFour$ sYM amplitudes should
have extensions away from four dimensions.  Motivated by this,
we write down generators for dual conformal transformations in six
dimensions and propose transformation properties, which are
straightforward to check at four points using the known expressions.

Finally, we also note that with the six-dimensional helicity formalism
contributions to amplitudes factorize into products of
chiral-conjugate pairs.  Since it holds in all cuts which decompose
loop amplitudes into sums of products of tree amplitudes, it seems
reasonable that it holds at loop level as well.  Although this
structure is somewhat reminiscent of the recently discovered
double-copy structure of gravity-diagram
numerators~\cite{BCJ,LoopBCJ}, as we shall see, there are a number of
significant differences.  Nevertheless, it can help simplify
calculations.

This paper is organized as follows: In
\sect{SixDimHelicityReviewSection} we review the six-dimensional
helicity formalism, showing the connection to four-dimensional
helicity, briefly review the unitarity method in the context of this
formalism, and discuss modifications needed for obtaining massive
regulators.  Then in \sect{DoubleCopySection} we review BCFW recursion
in six dimensions, before describing the chiral-conjugate property.
In \sect{OneLoopQCDSection} we turn to our first examples, which are
one-loop four-point amplitudes of QCD. In \sect{SuperHelicitySection}
we review an on-shell superspace compatible with six-dimensional
helicity. We give special care to the three-point amplitude since new
variables are needed. We also develop the necessary building blocks
for high-loop unitarity cuts using on-shell recursion. Then in
\sect{MultiLoopApplicationSection} we turn to high-loop applications,
using the two-loop four-point amplitude as an example to illustrate
the methods before turning to higher loops.  In
\sect{DualConformalSection} we then discuss an extension of dual
conformal properties to six dimensions. We give our comments and
prospects for future developments in \sect{ConclusionSection}.  We
include two appendices, one showing the analytic results of a
nontrivial cut of the four-loop four-point amplitude and the other
collecting auxiliary variables~\cite{CheungOConnell} needed for
defining three-point tree amplitudes.

\section{Six-dimensional helicity}
\label{SixDimHelicityReviewSection}

We begin by summarizing Cheung and O'Connell's six-dimensional
spinor-helicity formalism~\cite{CheungOConnell}.  To make the
connection to four-dimensional spinor techniques more transparent, we
choose to write the six-dimensional spinors in terms of the more
conventional four-dimensional ones~\cite{Boels,DHS}.  This form is
particularly well suited for carrying out loop calculations in
dimensional regularization, as needed for QCD computations.  It also
allows us to interpret the extra-dimensional contributions in terms of
masses.

\subsection{Spinor helicity}
\label{SixDimHelicitySubSection}

Before we review the six-dimensional helicity formalism, we state our
conventions of spinors in four dimensions. We follow ref.~\cite{TreeReview}
and work in a Weyl basis for the spinors such that
$p_{\alpha\dot\alpha}=\lambda_{p\alpha}\tilde{\lambda}_{p\dot\alpha}$.
We use the bra-ket notation for the chiral spinors and their
contractions,
\begin{eqnarray}
  \spa i.j=\epsilon^{\alpha\beta}\lambda_{i\,\beta}\lambda_{j\,\alpha}\,,\quad 
 \spb i.j=\epsilon_{\dot\alpha\dot\beta}\tilde\lambda_i^{\dot\beta}\tilde\lambda_j^{\dot\alpha}\,,\quad\mbox{and}\quad 
 \spa i.j\spb j.i=2\, p_i\!\cdot\!p_j = s_{ij}\,,
\end{eqnarray}
with $\epsilon_{12}=-1$, $\epsilon^{12}=1$. For further details we
refer to the above-mentioned
review.\footnote{Ref.~\cite{CheungOConnell} differ by a sign in 
  the definitions of four-dimensional spinor angle brackets 
  so that instead $\spa i.j\spb j.i =-s_{ij}$. 
  }

In six dimensions, a vector can be expressed using the spinor representation
of SO(5,1), 
\begin{equation}
p_{AB} =p_{\mu}\sigma^{\mu}_{AB}\,, \hskip 2 cm 
p^{AB} =p_{\mu} \tilde \sigma^{\mu AB}\,,
\end{equation}
where {\small $\{A,B\,\cdots \}$} are fundamental representation
indices of the covering group, SU$^*$(4).
Here the $\sigma^\mu_{AB}$ and $\tilde \sigma^{\mu
  AB}$ are antisymmetric 4$\times$4 matrices which play a role
analogous to the Pauli matrices in four dimensions.  Further details
and explicit forms of the matrices may be found in Appendix A of
ref.~\cite{CheungOConnell}.  The Dirac equation for Weyl spinors in
six dimensions can be written as,
\begin{equation}
	p_{\mu} \sigma^\mu_{AB} \lambda^{Ba}_p = 0\,, \hskip 2 cm 
	p_{\mu} \tilde\sigma^{\mu AB}\, \tilde\lambda_{pB\da} = 0\,,
\end{equation}
and gives rise to two independent solutions for each of the Weyl
spinors, $\lambda^{Ba}$ and $\tilde\lambda_{B\da}$. Each solution is
labeled by indices $a$ or $\dot a$, which are spinor indices of the
little group SO(4), corresponding to SU(2)$\times$SU(2).  We may lower
and raise the little group SU(2) indices $a,\dot a$ with the matrices
$\e_{ab}$ and $\e^{\da\db}$,
\begin{equation}
\lambda_a
=\e_{ab}\lambda^b\,,\hskip 2cm \tilde\lambda^\da
=\e^{\da\db}\tilde\lambda_\db\,,
\end{equation}
and $\e_{12}=-1$, $\e^{12}=1$, for this case as well.
Spinor inner products are defined by contractions of the SU$^*$(4) indices,
\begin{equation}
\langle i^a | j_\db ] = \lambda_i^{A a} \tlambda_{j A \db} = 
 [ j_\db | i^a\rangle\,.
\end{equation}
Other common quantities are spinors contracted with the SU$^*$(4)-invariant 
Levi-Civita tensor,
\begin{eqnarray}
\langle i^a j^b k^c l^d \rangle
&\equiv& \eps_{ABCD} \lambda_i^{Aa}\lambda_j^{Bb} \lambda_k^{Cc}
   \lambda_l^{Dd} \,, \nn\\
{}[ i_{\vph\da} j_\db k_{\vph\dc} l_\dd ]
&\equiv&
 \eps^{ABCD} \tlambda_{iA\da}\tlambda_{jB\db}\tlambda_{kC\dc}
   \tlambda_{lD\dd}\;.
\label{SpinorContractions}
\end{eqnarray}

The on-shell massless condition can be solved using bosonic
six-dimensional chiral spinors in a way similar to the well-known
four-dimensional case. The antisymmetry of $p_{AB}$ together with the
on-shell condition $p^2 \sim \e^{ABCD} p_{AB} p_{CD}= \det(\s p)=0$
gives the bispinor representation,
\begin{eqnarray} 
 p^{AB}=\lambda^{A a}\,\epsilon_{ab}\lambda^{B b} \,, \hskip 1 cm 
  p_{AB}=\tilde\lambda_{A\da}\,\epsilon^{\da\db}\tilde\lambda_{B\db} \,,
 \hskip 1 cm \lambda^{A a}\tilde{\lambda}_{A\dot{a}}=0 \,.
\end{eqnarray}
One can also express the vector momenta directly in terms 
of the spinors via,
\begin{equation}
p^\mu = - \frac{1}{4} \langle p^a |\sigma^\mu |p^b \rangle \eps_{ab} 
= -\frac{1}{4} [ p_\da | \tsigma^\mu | p_\db ] \eps^{\da \db}\,.
\end{equation}
Other useful quantities appearing in amplitudes are spinor strings,
\begin{eqnarray} 
\langle i^a| \s p_1 \s p_2 \cdots \s p_{2n +1} | j^b\rangle &=& {
(\lambda_i)^{A_1a}\, (p_1)_{A_1 A_2}\, (p_{2})^{A_2 A_3}
\cdots (p_{2n+1})_{A_{2n+1} A_{2n+2}}\, (\lambda_j)^{A_{2n+2}b}} \,,
\nn \\
\langle i^a| \s p_1 \s p_2 \cdots \s p_{2n} | j_\db] &=& {
(\lambda_i)^{A_1a}\, (p_1)_{A_1 A_2} \, (p_2)^{A_2 A_3}\, 
\cdots (p_{2n})^{A_{2n} A_{2n+1}}\, (\tilde\lambda_j)_{A_{2n+1}\db}} \,.
\label{SpinorStrings}
\end{eqnarray}

In six dimensions we have four polarization states. Following Cheung
and O'Connell, the polarization vectors can be written
as~\cite{CheungOConnell},
\begin{eqnarray}
  \pol^\mu_{a \da}(p,k) = \frac{1}{\sqrt{2}} 
              \langle p_a | \sigma^\mu|k_b\rangle 
             (\langle k_b | p^{\da}])^{-1}
                      =  
             (\langle p^a | k_{\db}])^{-1}
		     \frac{1}{\sqrt{2}} [ k_\db | \tsigma^\mu|p_\da ] 
	      \,,
\label{polar}
\end{eqnarray}
where the $a$ and $\da$ indices are labels for the four polarization
states.  As for the four-dimensional case, we have a null reference
momentum $k$ to define the states.  For each of the Weyl spinors
$[i^\da|$ and $\langle j_a|$ the indices $a$ and $\da$ label two
  helicity states respectively.  The object $(\langle p^a |
  k_{\db}])^{-1}=-(\langle p_a | k^{\db}])/2 p\cdot k$ is the
  inverse matrix of the spinor product $\langle p^a | k_{\db}]$ with
    respect to the helicity indices.  As for the four-dimensional case,
    the polarization vectors are transverse, and a redefinition of the
    reference spinor can be shown to correspond to a gauge
    transformation.

Given the proliferation of indices, we now summarize the indices that appear.
For four-dimensional objects the indices are,
\begin{eqnarray}
\hbox{SL(2,C) fundamental labels:}
  \quad && \alpha, \beta, \gamma, \ldots = 1,2\,,
       \hskip .3 cm \hbox{and} \hskip .3 cm
        \dot \alpha, \dot\beta, \dot\gamma, \ldots = 1,2\,, \nn \\
\hbox{SO(3,1) vector labels:}
  \quad && \mu, \nu, \rho, \ldots = 0,1,2,3 \,,
\end{eqnarray}
For six-dimensional objects the indices are,
\begin{eqnarray}
\hbox{SU$^*$(4) fundamental labels:}\quad && A,B,C, \ldots = 1,2,3,4 \nn\\
\hbox{SO(5,1) vector labels:}\quad && \mu, \nu, \rho, \ldots = 0,1,2,\dots5\, \nn\\
\hbox{SU(2)$\times$SU(2) helicity labels:}\quad &&  a,b,c, \ldots =1,2 \,
  \hskip .3 cm \hbox{and} \hskip .3 cm \da, \db, \dc, \ldots = 1,2 \,.
\end{eqnarray}
The six-dimensional spinors will be identifiable because they carry 
SU(2)$\times$SU(2) little-group helicity indices.

\subsection{Decomposing six-dimensional spinors into four-dimensional ones}
\label{FourDimHelicitySubSection}
 
For many purposes, it is convenient to express the six-dimensional
spinors in terms of the better-known four-dimensional ones~\cite{Boels,DHS}.
This allows us to define a physical four-dimensional subspace, and identify
states polarized within and transverse to it.

Viewed from four dimensions, the six-dimensional massless
condition is that of a massive four-dimensional vector,
\begin{equation}
  p^2 = \pfour^2 - p_4^2-p_5^2 \equiv \pfour^2 - m \tm = 0\,,  
\end{equation}
 where $\pfour$ denotes a
momentum vector keeping only the first four components.
The masses are related to the fifth and sixth components 
of the momentum,
\begin{equation}
m \equiv p_5 - i p_4\,, \hskip 1 cm \tilde m \equiv p_5 + i p_4\,,
\label{MassDef}
\end{equation}
where we have chosen the masses to be complex, since these will
be the natural combinations appearing in six-dimensional spinors
when decomposed into four-dimensional ones.
In general, massive momenta can be expressed in terms of two pairs
of four-dimensional spinors, here denoted by $\lambda, \tilde \lambda$ and 
$\mu, \tilde \mu$, via,
\begin{eqnarray}
&&\pfour_{\alpha\dot\alpha} = \lambda_\alpha\tilde\lambda_{\dot\alpha}+\rho\,\mu_\alpha\tilde\mu_{\dot\alpha} \,,
\label{MomentumFourDecomposition}
\end{eqnarray}
where, 
\begin{eqnarray}
	&&\quad\rho=\kappa\tilde\kappa=\kappa' \tilde\kappa' \,, \hskip .5 cm 
          \kappa\equiv\frac{m}{\spa{\lambda}.{\mu}}\,, \hskip .5 cm 
	  \tilde\kappa=\frac{\tm}{\spb{\mu}.{\lambda}}\,, \hskip .5 cm 
	  \kappa'\equiv\frac{\tm}{\spa{\lambda}.{\mu}}\,, \hskip .5 cm 
	  \tilde\kappa'=\frac{ m}{\spb{\mu}.{\lambda}}\,. \hskip .5 cm 
\label{MomentumFourConstraint}
\end{eqnarray}
Treating the six-dimensional spinors $\lambda^A{}_a$ and
$\tlambda{}_{A\dot a}$ as $4\times 2$ matrices, the six-dimensional massless
spinors can then be expressed in terms of the above massless
four-dimensional spinors via,
\begin{eqnarray}
&&	\lambda^A{}_a=\left(\begin{array}{cc}
			-\kappa\mu_\alpha&\lambda_\alpha\\ 
			\tilde\lambda^{\dot\alpha}&\tilde\kappa\tilde\mu^{\dot\alpha}
		\end{array}\right)\,,\qquad 
	\tilde \lambda_{A\dot a}=\left(\begin{array}{cc}
		\kappa'\mu^\alpha  &\lambda^\alpha\\ 
		-\tilde\lambda_{\dot\alpha}&\tilde\kappa'\tilde\mu_{\dot\alpha} 
		\end{array}\right)\,.
\label{SixDimSpinors}
\end{eqnarray}
The two columns in $\lambda^A{}_a$ and $\tlambda{}_{A\dot a}$ correspond to the little-group index $a$ and $\dot a$ taking value $1$ or $2$ respectively.   
This particular embedding of the four-dimensional spinors is constrained by the 
explicit form of the $\sigma^\mu_{AB}$ matrices used here. 
Using this, we can express the six-dimensional momenta in terms of the
antisymmetric matrices,
\begin{eqnarray}
&&	 p^{AB}=
	\left(\begin{array}{cc}
			- m\,\epsilon_{\alpha\beta}
			&\lambda_\alpha\tilde\lambda^{\dot\beta}+\rho\,\mu_\alpha\tilde\mu^{\dot\beta}    \\
			-\tilde\lambda^{\dot\alpha}\lambda_\beta-\rho\,\tilde\mu^{\dot\alpha}\mu_\beta &
			 \tm\,\epsilon^{\dot\alpha\dot\beta}
		\end{array}\right)\,, \nn\\
&&	p_{AB}=
	\left(\begin{array}{cc}
			 \tm\,\epsilon^{\alpha\beta}&
			\lambda^\alpha\tilde\lambda_{\dot\beta}+\rho\,\mu^\alpha\tilde\mu_{\dot\beta}\\
			-\tilde\lambda_{\dot\alpha}\lambda^\beta-\rho\,\tilde\mu_{\dot\alpha}\mu^\beta
			&- m\,\epsilon_{\dot\alpha\dot\beta}
		\end{array}\right)\,,
\end{eqnarray}
where $AB$ are the antisymmetric SU$^*$(4) indices.  

If we choose momenta which lie in the four-dimensional subspace
$p^{4,5}=0$, the spinors simplify to,
\begin{eqnarray}
&&	\lambda^A{}_a=\left(\begin{array}{cc}
			0 &\lambda_\alpha\\ 
			\tilde\lambda^{\dot\alpha}& 0
		\end{array}\right)\,, \hskip 1.5 cm 
	\tilde \lambda_{A\dot a}=\left(\begin{array}{cc}
		0  &\lambda^\alpha\\ 
		-\tilde\lambda_{\dot\alpha}& 0
		\end{array}\right)\,,\quad
		\label{6dspinors}
\end{eqnarray}
so that the six-dimensional spinors reduce to four-dimensional
spinors.

We may also relate the six-dimensional polarization vectors
(\ref{polar}) to four-dimensional ones. With four-dimensional
reference spinors embedded as the above momenta, the states with
$(a,\da)$ either $(1,\dot1)$ or $(2,\dot2)$ correspond to positive and
negative helicity states respectively. The mixed labels $(1,\dot2)$
and $(2,\dot1)$ appear as scalars from the four-dimensional point of
view.  Of course, the particular map between six-dimensional quantum
numbers and helicities depends on the explicit embedding of
four-dimensional spinors in the six-dimensional space.

\subsection{Tree-amplitude examples}
\label{TreeAmplitudeSubSection}

As usual, in this paper we will work with color-ordered
amplitudes~\cite{TreeReview}.  Applying the above helicity formul\ae{}  to the 
color-ordered four-gluon amplitude, one obtains a remarkably compact
result~\cite{CheungOConnell},
\begin{equation}
A^{\tree}_{4}(1^g_{\vph a \da}, 2^g_{b \db}, 3^g_{\vph c \dc}, 4^g_{\vph d \dd}) 
 = - \frac{i}{s_{12} s_{23}} 
\langle 1_a 2 _b 3_c 4_d \rangle [ 1_{\vph\da} 2 _\db 3_{\vph\dc} 4_\dd ]\,,
\label{FourGluonTree}
\end{equation}
where the inner products are defined in \eqn{SpinorContractions}.
As another example, the
two-chiral-quark two-gluon amplitude is just as simple and given
by,
\begin{equation}
 A^{\tree}_4(1^g_{\vph a \da},2^g_{b\db},3^q_{\vph c},4^q_{\vph d}) =
-\frac{i}{s_{12} s_{23}} \langle 1_a 2_b 3_c 4_d \rangle 
      [1_{\da} 2_{\db} 3_{\vph\dot{e}} 3^{\dot{e}}] \,,
\label{TwoQuarkTwoGluonTree}
\end{equation}
where the repeated index ``$\dot{e}$'' on the right side is summed.
Besides their remarkable simplicity, these formul\ae{} exhibit a
number of rather nice properties.  In six dimensions, because the
little group SU(2)$\times$SU(2) connects all helicities together, there
is no concept like maximally helicity violating (MHV) amplitudes: in
terms of six-dimensional spinors a single formula describes the
amplitude.  This property is manifest in \eqn{FourGluonTree}.

The three-point amplitude is a bit trickier.
Naively, the amplitude should contain an inverse of $\langle i_a |
j_\db]$ which would be ill-defined because $\det \langle i_a | j_\db]$
vanishes for three-point kinematics.  The solution worked out by
Cheung and O'Connell is to introduce two-component 
objects $u_{ia}, \tilde u_{j \db}$ and  $w_{ia}, \tilde w_{j \db}$.
In terms of these, the three-point amplitude is given by,
\begin{equation}
A_3^\tree (1_{\vph a \da}, 2_{b \db}, 3_{\vph c \dc}) = 
i \Gamma_{abc} \, \tilde{\Gamma}_{\da \db\dc}\,,
\label{ThreePointAmplitude}
\end{equation}
where,
\begin{eqnarray}
\Gamma_{abc} &=& u_{1a}  u_{2b} w_{3 c} + u_{1a} w_{2b} u_{3c} 
                + w_{1a} u_{2b} u_{3c} \,, \nn\\
\tilde{\Gamma}_{\da \db \dc} &=& \tilde{u}_{1\da}  \tilde{u}_{2\db} \tilde{w}_{3 \dc} 
                     + \tilde{u}_{1\da} \tilde{w}_{2\db} \tilde{u}_{3\dc} 
                     + \tilde{w}_{1\da} \tilde{u}_{2\db} \tilde{u}_{3\dc} \,.
\end{eqnarray}
The definitions of the $u$ and $w$ objects may be found in
ref.~\cite{CheungOConnell} and are summarized here 
in \app{ThreePointAppendix}. 

A striking property that can be observed in the three- and four-point
amplitudes (\ref{FourGluonTree}) and (\ref{ThreePointAmplitude}) is that
they contain pairs of chiral-conjugate factors.
 In particular, for the
four-gluon case, both chiral $\langle 1_a 2 _b 3_c 4_d \rangle$ and
antichiral $[ 1_{\vph\da} 2 _\db 3_{\vph \dc} 4_\dd ]$ factors appear
in the numerator. Similarly, the fermionic amplitude
(\ref{TwoQuarkTwoGluonTree}) breaks up into a chiral and an antichiral
factor. 

\subsection{Unitarity method and six-dimensional helicity}

The modern unitarity method constructs loop amplitudes directly from
on-shell tree amplitudes~\cite{UnitarityMethod}, by combining
unitarity cuts into complete expressions for amplitudes.  The most
convenient cuts generally are those that reduce the loop amplitude
integrands into a sum of products of tree amplitudes,
\begin{equation}
	\left.{A}^{\rm loop}_n\right|_{\rm cut}=
	\sum_{\rm states}
 		A^\tree A^\tree \cdots A^\tree A^\tree\,.
\label{Cuts}
\end{equation}
To construct the full amplitude, we can apply a merging procedure for
combining cuts~\cite{TwoLoopSplit}. For more complicated cases, it
is best to build an ansatz first in terms of some arbitrary
parameters.  The arbitrary parameters are then determined by comparing
the cuts of the ansatz against the generalized unitarity cuts
(\ref{Cuts}).  If an inconsistency is found, the ansatz is not general
enough and must be enlarged.  In four dimensions, a particularly simple
set of cuts to evaluate is those with the maximal numbers of cut
propagators~\cite{BCFUnitarity}.  Often it is convenient to build the
ansatz for the amplitude by starting with generalizations of these
types of maximal cuts and then systematically relaxing the cut
conditions one at a time~\cite{FiveLoop,CompactThree,FourLoop}.  This
approach, known as the method of maximal cuts, offers a systematic 
procedure for obtaining complete amplitudes,
including nonplanar contributions, at any loop order in massless
theories.  One can carry out cut calculations either analytically or
numerically comparing to a target ansatz at high precision.\footnote{
Since this procedure does not involve any integration, high precision
is straightforward.  One can also choose rational numbers for the
kinematic points.}

An important concept is ``spanning cuts'', or a complete set of cuts
for determining an amplitude.  For the color-ordered one-loop
four-point amplitude a spanning set is the usual $s$ and $t$ channel
two-particle cuts.  More generally, spanning cuts are determined by
requiring that all potential terms that can contribute, including
contact terms, can be detected by the cuts.  In
\sect{MultiLoopApplicationSection} we use such a spanning set of cuts
to confirm the six-dimensional validity of the complete four-loop
four-point amplitude of $\NeqFour$ sYM theory, calculated
in refs.~\cite{FourLoop,GravityFour} using mostly four-dimensional
methods.

One of the key ingredients in evaluating unitarity cuts is to have
kinematics that satisfies all cut conditions.  At one loop, for
explanatory purposes it is much more convenient to use analytic
solutions of the cut conditions, as we do in \sect{OneLoopQCDSection}, 
while at higher loops, numerical solutions are generally easier to use.  
One-loop analytic solutions for the on-shell conditions for massless 
and massive cases have been discussed previously in
refs.~\cite{BCFUnitarity,BrittoDDim,OPP,Kilgore,Forde,Badger}.

\subsection{Comments on connection to massive regulators}

For $\NeqFour$ sYM theory in four dimensions, one can use a massive
regulator, because the theory is only infrared divergent and not ultraviolet
divergent.  For planar amplitudes, a convenient massive Higgs regulator
was recently proposed in ref.~\cite{HiggsReg} for use in this
theory. This regulator has the advantage of preserving the dual
conformal invariance and also of leading to integrals that are much
simpler to evaluate than their dimensionally regularized counterparts.

In general, we can instead view a massively regulated amplitude as a
higher-dimensional one, but where we do not carry out the loop
momentum integrations over the extra-dimensional components.  This
meshes well with our six-dimensional helicity implementation of the
unitarity method.  The gauge invariance of this construction is
guaranteed by the gauge invariance of the constituent tree
amplitudes appearing in the unitarity cuts.\footnote{In theories with fewer
  supersymmetries, gauge invariance at the level of the integrand is
  complicated by the potential presence of terms such as bubbles on
  external legs, which are not detectable in the unitarity cuts.}  
To implement a massive regulator, we simply
interpret the extra components in terms of masses, $
\sum_{\mu = 4}^5 l_i^\mu l_{i\mu} \equiv - m_i^2$, taking on definite
values that are not integrated. The key constraints on the allowed
choices are that the underlying extra-dimensional components must
satisfy the usual momentum conservation constraints and that all
particles remain massless from the six-dimensional viewpoint.  From
the four-dimensional viewpoint, the on-shell condition is then, $
\sum_{\mu = 0}^3 l^\mu_i l_{i\mu} \equiv m_i^2$.
As emphasized in ref.~\cite{HiggsReg}, one particularly effective
choice for planar amplitudes is to take a uniform mass in the
outermost loop, while keeping all other legs massless. 

In many cases, conversion of massless dimensionally regularized
results to massive ones is straightforward~\cite{HiggsReg}.  However,
in general, one would like to derive or confirm massively regulated
integrands directly from unitarity cuts.  In particular, our
confirmation in \sect{FourloopFourpointSubsection} that the functional
form of the four-point four-loop amplitude is unchanged between four
and six dimensions ensures that the straightforward conversion of this
amplitude to a massively regulated form is compatible with unitarity.
For six- and higher-point amplitudes, we know that the situation is a
bit more complicated, because the higher-dimensional integrands
contain additional pieces not detectable in four-dimensional massless
cuts.  Although such terms are expected to vanish as $m \rightarrow 0$
even after integration, they would be needed for studies of the
Coulomb phase of the theory where the mass is kept finite.


\section{A six-dimensional structure}
\label{DoubleCopySection}

As seen in \sect{TreeAmplitudeSubSection}, three- and four-point tree
amplitudes contain a pair of chiral-conjugate factors.  How general is
this?  To answer this we turn to BCFW recursion relations~\cite{BCFW},
which offer a means for constructing tree amplitudes in a form where
we can exploit the six-dimensional helicity techniques.  It also
provides a convenient means for generating tree amplitudes needed in
the unitarity cuts.

\subsection{BCFW recursion}

The BCFW shift using six-dimensional spinor helicity was given in
ref.~\cite{CheungOConnell}, and the supersymmetric version
in~\cite{DHS}.  Here we give a brief summary and the relevant bosonic
shifts. We begin by picking two external lines, say $1,2$ as special,
and deform them by a null vector proportional to a complex parameter
$z$,
\begin{equation}
{p}_{1}(z) =p_1+z \shift, \hskip 1 cm \,p_{2}(z)=p_2-z\shift \,,
\label{MomentumShift}
\end{equation}
where $\shift$ is a null vector satisfying $p_1\cdot \shift=p_2 \cdot
\shift=0$.  These conditions ensure that the deformed momenta remain
on-shell, $p^2_1(z)= p^2_2(z)=0$, and the overall momentum
conservation of the amplitude is unaltered.

Since a tree amplitude is a rational function of momenta with no more
than a single pole in any given kinematic invariant, this
deformation will result in a complex function with only simple
poles in $z$. Each pole in $z$ will correspond 
to a propagator that is a sum of a subset
of external momenta that includes only one of the shifted momenta,
\begin{equation}
P_{2j}(z)^2\equiv (p_2(z)+\ldots+ p_j)^2= P^2_{2j} + 2 z \shift\cdot P_{2j}\,.
\end{equation}
The location of the pole is given by solving 
the on-shell condition $P_{2j}(z)^2 = 0$,
\begin{equation}
z_{2j} = -\frac{P^2_{2j}}{2\shift\cdot P_{2j}} \,.
\label{FrozenZ}
\end{equation}
As in common usage, we denote the shifted momenta evaluated at the value
of $z$ locating a pole with a hatted symbol ``$\hat{\quad}$'', e.g.
$\hat p_2 \equiv p_2(z = z_{2j})$.

If a shifted amplitude vanishes as $z\rightarrow \infty$, then
standard complex variable theory implies that it is uniquely
determined by its residues in $z$. For $D\ge 4$ dimensions, it is
straightforward to find choices of shifts where this property holds for
Yang-Mills theories~\cite{NimaDDim,CheungOConnell}.  The poles
correspond to configurations where propagators go on shell and where the
amplitude factorizes into a product of lower-point amplitudes. Each
residue is then simply a sum of products of two lower-point tree
amplitudes on either side of the propagator, with the shifted momenta
evaluated at the location of the pole. If legs 1 and 2
are shifted, the BCFW recursion relation gives us the unshifted
amplitude as,
\begin{equation}
A_n^\tree(0)=\sum_{j=3}^{n-1}\sum_{h}
  A_L\Bigl(\hat{p}_2,\ldots,p_j,-\hat{P}^{(-h)}_{2j}\Bigr) 
  \frac{i}{P_{2j}^2}
  A_R\Bigl(\hat{P}^{(h)}_{2j},p_{j+1},\ldots, \hat{p}_1\Bigr)
  \biggl|_{z=z_{2j}} \,,
\label{1stBCFW} 
\end{equation}
where $A_L$ and $A_R$ are lower-point tree amplitudes on the left and right
sides of the propagator and are evaluated at shifted momenta with 
$z =z_{2j}$.
The first sum in the above equation is over all diagrams that produce
poles, while the second is over all helicity states that cross the
internal leg $\hat P$.

In six dimensions, the conditions $p_1\cdot \shift=p_2 \cdot \shift=0$ can be
solved by choosing $\shift$ to be proportional to the polarization of line
1, $\shift\sim\e_{1a\dot{a}}$, and choosing the reference spinor for the
polarization vector to be $\lambda_2$.  This 
satisfies the needed constraints and leads to sufficiently good
behavior at large $z$.  Since $\e_{1a\dot{a}}$ has extra
SU(2)$\times$SU(2) little-group indices, one introduces an auxiliary
matrix $X^{a\dot{a}}$ to contract the indices so that,
\begin{equation}
\shift^{AB}=\frac{1}{\sqrt{2}}X^{a\da}(\epsilon^{AB}_1)_{a\da}
= - X^{a\da}\frac{\lambda^{[A}_{1a}\lambda^{B]}_{2b}}{[1^{\da}|2_b\rangle}
= X^{a\da}\frac{^{[A}\s{p}_2|1_{\da}]\langle1_a|^{B]}}{s_{12}} \,,
\end{equation}
and the condition $\shift^2=0$ now becomes $\det X=0$.  This can
automatically be satisfied by choosing $X_{a\dot{a}}=x_a
\tilde{x}_{\da}$, where the $x_a$ and $\tilde{x}_{\da}$ are arbitrary.
These arbitrary variables will cancel out once one sums over all
contributions. When implementing the recursion relations numerically,
these variables are helpful since they can identify errors when one checks
whether the result is independent of the choice of $x_a,\;
\tilde{x}_{\dot{a}}$.  The above shift can be translated into a
redefinition of the spinors,
\begin{eqnarray}
\lambda^{Aa}_{{1}}(z)&=&\lambda^{Aa}_{1}
  - \frac{z}{s_{12}}\, X^a_{\;\;\da} [1^{\da}|2^b\rangle\lambda^A_{2b}  \,,\hskip 1.3 cm
 \lambda^{Ab}_{{2}}(z)=\lambda^{Ab}_{2}
  - \frac{z}{s_{12}}\, X^{a}_{\;\;\da} \lambda^A_{1a}[1^{\da}|2^b\rangle \,, \nn \\
\tilde{\lambda}_{{1}A\da}(z)&=&\tilde{\lambda}_{1A\da}
 +\frac{z}{s_{12}}\, X^a_{\;\;\da} \langle 1_{a}|2_{\db}]\tilde{\lambda}_{2A}^{\db}\,, \hskip 0.95 cm 
\tilde{\lambda}_{{2}A\db}(z)=\tilde{\lambda}_{2A\db}
  + \frac{z}{s_{12}}\, X^{a}_{\;\;\da} \tilde{\lambda}_{1A}^{\da}\langle 1_a|2_{\db}]\,. \hskip 1 cm 
\label{bshift}
\end{eqnarray}

In \sect{SuperBCFWSubsection}, we will describe the supersymmetrized
BCFW recursion relations to generate the tree amplitudes needed in
calculations of multiloop amplitudes of $\NeqFour$ sYM
theory.

\subsection{A double copy}
\label{DoubleCopySubsection}

As noted in \sect{TreeAmplitudeSubSection}, the three- and four-point
amplitudes contain chiral-conjugate factors.  An obvious question is
whether this property can be extended to $n$ points. Indeed this is
the case, as can be seen using BCFW recursion relations.  This can be
understood by noting that six-dimensional on-shell helicity states of
Yang-Mills theory transform under a representation of the little group
which factorizes into the direct product of two spinor
representations. That is, the helicity sum in \eqn{1stBCFW}, can be
rewritten as a product of two independent sums: $\sum_{h}\rightarrow
\sum_{a}\times \sum_{\da}.$ Thus, if the lower-point amplitudes in the
numerator of \eqn{1stBCFW} can also be written as products, one factor
carrying the undotted SU(2) indices and the other carrying the dotted
SU(2) indices, then the BCFW recursion relations will preserve this
property for the higher-point amplitudes.  Since the three and
four-point amplitudes do have this property, it holds for any number
of legs by recursively constructing the amplitudes.

For example, at five points, applying a BCFW recursion
based on shifting legs 1 and 2, following the discussion of
ref.~\cite{CheungOConnell}, we obtain two contributions, one in
the  $s_{23}$ channel and the other in the $s_{51}$ channel.  Evaluating
the diagrams gives,
\begin{eqnarray}
\nonumber
&&\hskip -.5 cm 
 A_5^\tree(1_{a\da},2_{b\db},3_{c\dc},4_{d\dd},5_{e\de})\nn \\
&& \null \hskip .2 cm 
=-\frac{i}{s_{45} s_{5\hat 1}s_{23}}
\left(\langle\hat{1}_a\hat{2}_b4_d5_e\rangle u_{3c}
+\langle\hat{1}_a3_c4_d5_e\rangle u_{\hat 2b}\right)
\left(\left[\hat{1}_{\da}\hat{2}_{\db}4_{\dd}5_{\vph\dot{e}}\right]
\tilde{u}_{3\dc}+\left[\hat{1}_{\da}3_{\dc}4_{\dd}5_{\vph\dot{e}}\right]
\tilde{u}_{\hat2\dot{b}}
\right) \nn \\ 
&& \null \hskip .68 cm 
-\frac{i}{s_{34}s_{51}s_{\hat 2 3}}\left(\langle\hat{2}_b\hat{1}_a4_d3_c\rangle
u_{5e}+\langle\hat{2}_b5_e4_d3_c\rangle
u_{\hat 1a}\right)
\left(\left[\hat{2}_{\dot{b}}\hat{1}_{\da}4_{\dd}3_{\dc}\right]
\tilde{u}_{5\vph\dot{e}}+\left[\hat{2}_{\db}5_{\vph\dot{e}}4_{\dd}3_{\dc}\right]
\tilde{u}_{\hat 1 \da} \right) \,. \hskip 1 cm 
\label{FivePointTree}
\end{eqnarray}
We thus explicitly see that the chiral-conjugate nature of the
three- and four-point amplitudes is inherited directly by the five-point
amplitude via the BCFW recursion relations.  As noted above, the hats
indicate that the momenta are shifted as in \eqn{MomentumShift}, with
$z$ evaluated at the value (\ref{FrozenZ}) for the two given channels.  If we
try to clean up the expression, for example by factoring out the
auxiliary variables $x^a,\,\tilde{x}^{\dot{a}}$, we can easily hide the
 chiral-conjugate structure.  By continuing the recursion and maintaining the
BCFW format at each level of the recursion, we straightforwardly
obtain forms where chiral-conjugate factors appear
in each term at $n$ points.

This behavior has some similarities with the double-copy structure for
gravity amplitudes uncovered by two of the authors and Johansson (BCJ)
in ref.~\cite{BCJ}.  In that case, numerators of gravity diagrams
appear as two factors of gauge theory numerators.  There are, however,
also important differences between the two cases.  The most obvious
one is that in the gravity case no chiral conjugation is needed for
one of the copies.  More importantly, for the case of gravity, we can
interpret the factors as pieces of physical Yang-Mills amplitudes, but
for the present case of Yang-Mills theory we do not have an interpretation in
terms of amplitudes of some theory. In fact, it is unclear how to
write down a Lagrangian that would lead to only one of the factors of
the three-point Yang-Mills amplitude, either $\Gamma_{abc}$ or
$\tilde{\Gamma}_{\da\db\dc}$.  This may be contrasted with the gravity
case where there is a partial Lagrangian understanding~\cite{Square}.
Another difference is that the gravity double copy relies on a
group-theory Jacobi identity structure in the gauge-theory diagram
kinematic numerators~\cite{BCJ}. Such a structure is not apparent in
the present case.  Nevertheless, the present structure can be helpful
in computations.

\section{Example: One-loop QCD amplitudes}
\label{OneLoopQCDSection}

Our task in this section is to demonstrate how six-dimensional
helicity methods can be combined with the unitarity method to obtain
complete one-loop QCD amplitudes.  We take external momenta
and helicities to live in a four-dimensional subspace, but take the
internal momenta to be six dimensional.  In general, any
dimensionally regularized massless amplitude 
can be expressed as a linear
combination of scalar box, triangle, and bubble integrals with
rational coefficients~\cite{IntReduction}.  Since these integrals are
known, the problem is reduced to determining the rational coefficients
of the integrals.  Here we choose a slightly different representation
allowing also integrals with arbitrary powers of $m \tm$ in their
numerators, following refs.~\cite{BernMorgan,Badger}.

\subsection{One-loop four-point cut conditions in $D=6$}

\begin{figure}[th]
\centerline{\epsfxsize 5. truein \epsfbox{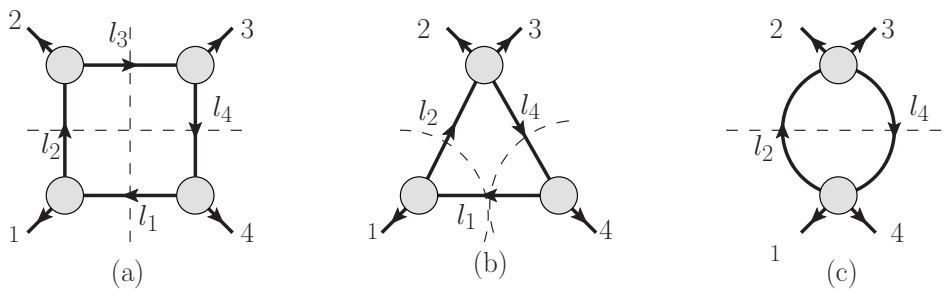}}
\caption[a]{\small Cuts used to obtain a one-loop four-point
  amplitude.  We display the triple cuts and ordinary two-particle cut
  in the $s_{14}$ channel. In general, we must also evaluate $s_{12}$
  channel cuts.}
\label{OneLoopCutsFigure}
\end{figure}

For the purposes of obtaining one-loop QCD amplitudes, we take the
external legs to carry massless four-dimensional momenta and the loop
to carry six-dimensional momentum.  To solve the cut conditions, we
view the massless six-dimensional loop momentum as massive
four-dimensional momentum.  The mass parameter is then conserved, as
inherited from the six-dimensional momentum components.  To solve for
the cut conditions, we consider a particle with a uniform mass around
the loop.

To illustrate six-dimensional helicity, we consider the simple case of
four massless external legs. 
We start with the quadruple cut displayed in
\fig{OneLoopCutsFigure}(a).  In this case, we must solve the cut
conditions $\lfour_i^2 = m \tm$, where $\lfour$ denotes a momentum
vector keeping only the first four components.  In this case we can
find a solution for the first four components of the loop momentum
vectors, in terms of purely four-dimensional spinors,
\begin{eqnarray}
(\lfour^\pm)_1^\mu &=& \frac{1}{2}
     \left(c_1^\pm \,\sand4.\gamma^\mu.1-\frac{1}{c_1^\pm}\,
            \frac{m \tm}{s_{14}} \sand1.{\gamma^\mu}.4\right) \,, 
  \hskip 1 cm \mu = 0,1,2,3  
\label{QuadCutSolution}
\end{eqnarray}
where, 
\begin{eqnarray} 
c^{\pm}_1=\frac{\spa1.2}{2\spa4.2}\left(1\pm \sqrt{1+\frac{4 m \tm s_{24}}{s_{23}s_{34}}}\right)\,.
\label{C1Solution}
\end{eqnarray}
Although there are two solutions to the quadruple cuts, 
in general, when summing over these the square root drops out.

The fifth and sixth dimensional components of the loop momenta are
\begin{equation}
l_i^4 = \frac{-m+\tm}{2i}\,,\qquad \qquad  l_i^5=\frac{m+\tm}{2} \,.
\label{QuadCutSolution6}
\end{equation}
The other momenta in the loop are given by momentum conservation $l_2=
l_1 - p_1$, $l_3 = l_1 - p_1 -p_2$ and $l_4 = l_1 + p_4$.  Because of
the symmetry in the problem we can also obtain these momenta from
relabellings of \eqn{QuadCutSolution}.  If viewed as massive momenta in
four dimensions, $m$ and $\tm$ are constants, and the momenta
(\ref{QuadCutSolution}) are two independent solutions to the cut
conditions.  In contrast, when interpreted as six-dimensional momenta,
(\ref{QuadCutSolution}) together with (\ref{QuadCutSolution6}) are not
constrained completely by the cut conditions.  Their remaining freedom
is parametrized by the now dynamic mass parameters $m$ and $\tm$. In
fact, what we labeled as two parameters can be thought of as a single
complex parameter family of solutions.  

Now consider the triple cut displayed in \fig{OneLoopCutsFigure}(b). 
We note that, 
\begin{eqnarray}
\lfour_1^\mu=  \frac{1}{2} \left( t \,\sand4.{\gamma^\mu}.1 
  - \frac{1}{t}\,\frac{m \tm}{s_{14}} \sand1.{\gamma^\mu}.4 \right)\,,
\label{TripleCutMomentum}
\end{eqnarray}
solves the three cut conditions $\lfour_1^2 = \lfour_2^2 = \lfour_4^2
= m \tm$, where $t$ is a parameter describing the remaining degree of
freedom (from the four-dimensional vantage point) not frozen by the
cut conditions.  The other two cut momenta are specified by momentum
conservation, $l_2 = l_1 -p_1$ and $l_4 = l_1 +p_4$. Again it is
simple to check that the cut legs all satisfy the proper
six-dimensional on-shell conditions.

We note that the similarity of the triple cut solution to the
quadruple cut solutions is not accidental.  Indeed if we demand that a
fourth propagator goes on shell, $\lfour_3^2 - m \tm = l_3^2 = (l_1-
p_1 - p_2)^2\rightarrow 0$, this has the effect converting the triple
cut into a quadruple cut.  That is, we demand,
\begin{equation}
0= (l_1 - p_1 - p_2)^2= 
 t \,\spa4.2\spb1.2+s_{12}-\frac{m \tm}{ t s_{14}}\spa1.2\spb4.2\,.
\end{equation}
Solving for $t$ gives,
\begin{equation}
t_0 = \frac{\spa1.2}{2\spa4.2}\left(1\pm\sqrt{1+\frac{
	4m \tm s_{24}}{s_{14}s_{12}}}\right)\,,
\label{QuadCutt0}
\end{equation}
matching $c_1^\pm$ in \eqn{C1Solution}.

\subsection{One-loop four-point solution of six-dimensional spinors}

Given the solution of the cut conditions above, it is then 
a simple matter to insert these into our solution for the 
six-dimensional spinors in terms of four-dimensional ones.
To illustrate this, consider the triple cut displayed in
\fig{OneLoopCutsFigure}(b).  To make the process simpler, it is
convenient to express the four-dimensional momentum solutions in
two-component notation.  Taking the solution (\ref{TripleCutMomentum}) and
rewriting it in two-component notation, we have,
\begin{eqnarray}
\lfour_1&=&t\lambda_4\tilde\lambda_1 -\frac{1}{t}\frac{m \tm}{s_{14}}\lambda_1\tilde\lambda_4 \,, \nn\\
\lfour_2&=&\left(\lambda_4
   -\frac{1}{t}\,\lambda_1 \right) t \tilde\lambda_1  
-
\frac{m \tm}{ t s_{14}}\lambda_1\,\tilde\lambda_4\,, \nn \\
\lfour_4&=&\lambda_4 \left(\tilde\lambda_4+ t\,\tilde\lambda_1\right)-
 \frac{m \tm}{ t s_{14}}\lambda_1\, \tilde\lambda_4 \,.
\label{TwoComponentSolution}
\end{eqnarray}
To arrange the six-dimensional spinors for the loop momentum in terms
of four-dimensional ones, we make use of the decomposition in
\eqns{MomentumFourDecomposition}{MomentumFourConstraint}.  Comparing
$l_1$ in \eqn{TwoComponentSolution} to \eqn{MomentumFourDecomposition},
we see that in this case,
\begin{eqnarray}
&& \mu=\lambda_1\,,\quad \tilde\mu= \frac{\tilde\lambda_4}{ t} \,,\quad 
    \lambda=\lambda_4\,,\quad \tilde\lambda= t\tilde\lambda_1\,,
   \nn\\
&&	\kappa_{14}=\frac{m}{\spa1.4}\,,\quad 
	\tilde\kappa_{14}=\frac{\tm}{\spb1.4}\,,\quad 
	\kappa_{14}'=\frac{\tm}{\spa1.4}\,,\quad 
	\tilde\kappa_{14}'=\frac{m}{\spb1.4}\,.
\end{eqnarray}
Plugging this into \eqn{SixDimSpinors} immediately gives us 
the six-dimensional spinors corresponding to $l_1$ as,
\begin{eqnarray}
&&	(\lambda_{l_1})^A{}_a=\left(\begin{array}{ll}
		-\kappa_{14}\lambda_1 & \quad \;
		\lambda_4\\
		t \tilde\lambda_1& \;\;
		\tilde\kappa_{14}\tilde\lambda_4/ t
	\end{array}\right)\,,\hskip 1 cm
	(\tlambda_{l_1})_{A\dot a}=\left(\begin{array}{ll}
		\quad \kappa_{14}'\lambda_1 & \quad \;
		\lambda_4\\
		- t \tilde\lambda_1& \;\;
		\tilde\kappa_{14}'\tilde\lambda_4/ t
	\end{array}\right)\,.  \hskip 1 cm 
\end{eqnarray}
Similarly, for the spinors carrying momenta $l_2$ and $l_4$ --- needed
for evaluating the triple cut in \fig{OneLoopCutsFigure}(b) ---
we have 
\begin{eqnarray}
&&	(\lambda_{l_4})^A{}_a=\left(\begin{array}{ll}
	\;\;   -\kappa_{14}\lambda_1 & \quad
		\lambda_4\\
		\tilde\lambda_4+ t \tilde\lambda_1&\;
		\tilde\kappa_{14}\tilde\lambda_4/t
	\end{array}\right)\,, \hskip 1.6 cm 
	(\tlambda_{l_4})_{A\dot a}=\left(\begin{array}{ll}
		\;\quad \kappa_{14}'\lambda_1 & \quad
		\lambda_4\\
		-\tilde\lambda_4 -t \tilde\lambda_1& \;
		\tilde\kappa_{14}'\tilde\lambda_4/t
	\end{array}\right)\,, \nn\\
&&	(\lambda_{l_2})^A{}_a=\left(\begin{array}{ll}
		-\kappa_{14}\lambda_1 &\;
		\lambda_4 - \lambda_1/t\\
		t\tilde\lambda_1& \quad
		\tilde\kappa_{14}\tilde\lambda_4/t
	\end{array}\right)\,, \hskip 1.4 cm 
	(\tlambda_{l_2})_{A\dot a}=\left(\begin{array}{ll}
		\kappa_{14}'\lambda_1 &
		\lambda_4- \lambda_1/t\\
		-t\tilde\lambda_1& \;\;
		\tilde\kappa_{14}'\tilde\lambda_4/t
	\end{array}\right)\,. \hskip 1.2 cm 
	\label{spinorcutsolution}
\end{eqnarray}

In many cases, we find ourselves with an incoming momentum, and as such
it is labelled $-p$ in our all-outgoing convention.
A simple prescription which preserves the relation between
spinors and momenta and which is sufficient for purely gluonic amplitudes
is to simply take $\lambda_{-p} = i \lambda_{p}$ and $\tlambda_{-p} = i
\tlambda_{p}$. A similar prescription for any state in the $\NeqFour$
supermultiplet will be given in \sect{SuperHelicitySection}.

Using this it is then a simple matter to work out the result for
the cut.  For the $s_{14}$-channel cut we have, if legs $1$ and $4$ are 
of positive helicity, 
\begin{eqnarray}
	&&\langle  ({-l_4})_1,4_1,1_1,(l_2)_1 \rangle
	=\det\left(\begin{array}{cccc}
	-i\kappa_{14}\lambda_1&0& 0 &-\kappa_{14}\lambda_1\\
	i\tilde \lambda_4 +  it \tilde\lambda_1&\, \tilde\lambda_4& 
          \, \tilde \lambda_1 &\,  t\tilde\lambda_1
	\end{array}\right)
	=0\,, \nn \\
	&&\langle  ({-l_4})_1,4_1,1_1,({l_2})_2 \rangle
	=\det\left(\begin{array}{cccc}
	-i\kappa_{14}\lambda_1&0&0 &\lambda_4- \lambda_1/t\\
	i\tilde \lambda_4 + i t\tilde\lambda_1 &\,\tilde\lambda_4& \,
        \tilde \lambda_1&\tilde\kappa_{14}\tilde\lambda_4/t
	\end{array}\right)
	= i\kappa_{14} \spa1.4 \spb4.1 \,, \nn \\
	&&\langle  ({-l_4})_2,4_1,1_1,({l_2})_1 \rangle
	=\det\left(\begin{array}{cccc}
	i\lambda_4&0&0&  - \kappa_{14}\lambda_1\\
	i\tilde\kappa_{14}\tilde\lambda_4/t& \, \tilde\lambda_4
         &\, \tilde \lambda_1& \, t \tilde\lambda_1
	\end{array}\right)
	=  -i \kappa_{14} \spa1.4  \spb4.1 \,, \nn \\
	&&\langle  ({-l_4})_2,4_1,1_1,({l_2})_2 \rangle
	=\det\left(\begin{array}{cccc}
	i\lambda_4&0& 0 &\lambda_4 - \lambda_1/t\\
	i\tilde\kappa_{14}\tilde\lambda_4/t&\,\tilde\lambda_4
         &\,\tilde \lambda_1& \,\tilde\kappa_{14}\tilde\lambda_4/t
	\end{array}\right)
	= - i\frac{\spa1.4 \spb4.1}{ t} \,.
\label{AllPlusSixD}
\end{eqnarray}
As seen from \eqns{SpinorContractions}{SixDimSpinors}, the self-conjugate  
structure implies that the bracket
expression is basically the same as the angle expression, except for 
the replacement $\kappa_{14}\rightarrow\kappa_{14}'$ along with some
signs, 
\begin{eqnarray}
	&& [ (-l_2)_{\dot 1}, 4_{\dot 1},1_{\dot 1},(l_4)_{\dot 1} ] = 0\,, 
        \hskip 3.2cm 
	   [ (-l_{2})_{\dot 1}, 4_{\dot 1},1_{\dot 1},(l_4)_{\dot 2}]
	= -i\kappa_{14}' \spa1.4 \spb4.1 \,, \nn \\
	&& [  (-l_2)_{\dot 2},4_{\dot 1},1_{\dot 1},(l_4)_{\dot 1} ]
	= i  \kappa_{14}' \spa1.4  \spb4.1 \,, \hskip 1 cm 
           [ ({-l_{ 2}})_{\dot 2},4_{\dot 1},1_{\dot 1},({l_4})_{\dot 2} ]
	= -i \frac{\spa1.4 \spb4.1}{ t} \,. \hskip 2 cm 
\label{AllPlusSixD2}
\end{eqnarray}
Similarly, if leg 1 is of negative helicity and leg 2 of positive, we
have,
\begin{eqnarray}
	&&\langle  ({-l_4})_1,4_1,1_2,({l_2})_1 \rangle
	=\det\left(\begin{array}{cccc}
	-i\kappa_{14}\lambda_1&0& \lambda_1 &-\kappa_{14}\lambda_1\\
	i\tilde \lambda_4 +  i t \tilde\lambda_1 &\,\tilde\lambda_4& 
           0 &t\tilde\lambda_1
	\end{array}\right)
	=0\,, \nn \\
	&&\langle  ({-l_4})_1,4_1,1_2,({l_2})_2 \rangle
	=\det\left(\begin{array}{cccc}
	-i\kappa_{14}\lambda_1&0& \lambda_1 &\lambda_4- \lambda_1/t\\
	i\tilde \lambda_4 + i t\tilde\lambda_1&\tilde\lambda_4&
         0 &\tilde\kappa_{14}\tilde\lambda_4/t
	\end{array}\right)
	= i\spa1.4 \spb4.1 t \,, \nn \\
	&&\langle  ({-l_4})_2,4_1,1_2,({l_2})_1 \rangle
	=\det\left(\begin{array}{cccc}
	i\lambda_4&0& \lambda_1&  - \kappa_{14}\lambda_1\\
	i \tilde\kappa_{14}\tilde\lambda_4/t&\tilde\lambda_4
         & 0 & t \tilde\lambda_1
	\end{array}\right)
	=  -i \spa1.4  \spb4.1 t\,, \nn \\
	&&\langle  ({-l_4})_2,4_1,1_2,({l_2})_2 \rangle
	=\det\left(\begin{array}{cccc}
	i \lambda_4&0& \lambda_1 &\lambda_4 - \lambda_1/t\\
	i \tilde\kappa_{14}\tilde\lambda_4/t&\tilde\lambda_4
         &0&\tilde\kappa_{14}\tilde\lambda_4/t
	\end{array}\right) = 0 \,,
\label{SingleMinusSixD}
\end{eqnarray}
and for the chiral conjugate,
\begin{eqnarray}
	&& [ ({-l_4})_{\dot{1}},4_{\dot{1}},1_{\dot{2}},({l_2})_{\dot{1}} ] = 0\,, \hskip 3.95cm 
	[ ({-l_4})_{\dot{1}},4_{\dot{1}},1_{\dot{2}},({l_2})_{\dot{2}} ]
	= i \spa1.4 \spb4.1 t \,, \nn \\
	&& [ ({-l_4})_{\dot{2}},4_{\dot{1}},1_{\dot{2}},({l_2})_{\dot{1}} ]
	= -i \spa1.4  \spb4.1 t\,, \hskip 1.8cm  
	[({-l_4})_{\dot{2}},4_{\dot{1}},1_{\dot{2}},({l_2})_{\dot{2}} ]	= 0 \,. \hskip 2.4 cm 
\label{SingleMinusSixDConj}
\end{eqnarray}
This solution is valid as well for quadruple cuts by substituting
$t \rightarrow t_0$ from \eqn{QuadCutt0}.

Similarly, the six-dimensional spinors can be solved for explicitly
for any one-loop cuts with any number of external legs, although we
refrain from giving the solutions here.  These solutions are closely
connected to the solutions of the cut conditions in four dimensions
with masses~\cite{BCFUnitarity,Kilgore,Badger}.

\subsection{QCD Calculations}
By treating the extra-dimensional components of loop momenta
effectively as masses to be integrated over, we can make use of
four-dimensional methods for evaluating the coefficients of the
integrals.  The box coefficients are obtained by using the four
on-shell conditions to freeze four components of loop
momenta~\cite{BCFUnitarity}.  For the triangle and bubble integrals, we can
use the approach of refs.~\cite{Forde,Badger}.

\begin{figure}[tb]
\centerline{\epsfxsize 1.7 truein \epsfbox{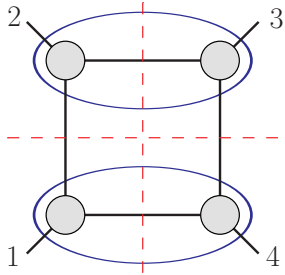}}
\caption[a]{\small The quadruple cut.  The cuts are indicated by
  dashed (red) lines.  Here two pairs of three-point amplitudes are
  grouped together to form two easier-to-use four-point tree
  amplitudes, indicated by the (blue) ovals. The cut propagators of
  the four-point tree amplitudes are canceled by multiplying by
  inverse propagators prior to imposing cut conditions.}
\label{QuadCutGroupFigure}
\end{figure}

As a warmup we consider the simplest one-loop QCD amplitudes, with
four identical-helicity external gluons in a four-dimensional
subspace.  This amplitude is especially simple because it can be
determined entirely from the quadruple cut illustrated in
\fig{QuadCutGroupFigure}, where all four propagators are placed on
shell.  This follows from the property that in both two-particle cuts
the four-point tree amplitudes on either side of the cut are each
proportional to $m \tm$.  Thus the numerator of the integrand must be
proportional to $(m \tm)^2$, saturating the dimensions of the
numerator polynomial.  Since this factor does not vanish when all four
box propagators are placed on shell, the quadruple cut is sufficient
for constructing the entire amplitude.

As discussed in \sect{TreeAmplitudeSubSection}, four-point tree
amplitudes expressed in terms of six-dimensional spinor helicity are
simpler than their three-point counterparts.  In general, this makes
it advantageous to group three-point amplitudes appearing in the cuts
into four-point amplitudes, but with the cut propagators removed by
multiplying by the appropriate inverse propagators.  With the
grouping illustrated in \fig{QuadCutGroupFigure} by the (blue) ovals,
we have the quadruple cut as,
\begin{eqnarray}
C_{1234} \!
&=& \!  \sum_{\rm states} (-i)^2 (l_2 - p_2)^2 (l_2 + p_1)^2\,
           A_4(-l_2, 2^+,3^+,l_4) \, A_4(-l_4, 4^+,1^+,l_2)  
 \label{QuadCutAllPlusNaive}   \\
&=& 
          \frac{ \langle (-l_2)_a, 2_1, 3_1, (l_4)_b \rangle \,
            [ (-l_2)_\da, 2_{\dot{1}}, 3_{\dot{1}}, (l_4)_\db ] \,
           \langle (-l_4)^a, 4_1, 1_1, (l_2)^b \rangle \,
            [ (-l_4)^\da, 4_{\dot{1}}, 1_{\dot{1}}, (l_2)^\db ] } 
           {s_{14} s_{23}}  \,.
 \nn
\end{eqnarray}
In general, we will label the
generalized cuts with the indices $i$ of $l_i$ labeling the
propagators; here all four propagators carrying momenta $l_1, l_2,
l_3$ and $l_4$ are cut.  We also assigned SU(2)$\times$SU(2) little
group indices $a,\da=1$ to the external legs so that they carry
positive helicity in the four-dimensional subspace.

Using the solution of the inner products given in \eqn{AllPlusSixD},
relabeled for the other side of the cut, the
quadruple cut in \eqn{QuadCutAllPlusNaive} 
gives,
\begin{eqnarray}
C_{1234} &=& 4
\frac{1}{s_{23} s_{14}} \kappa_{41} \kappa_{23} \kappa'_{41} \kappa'_{23}
\spa1.4^2 \spb4.1^2 \spa3.2^2 \spb2.3^2 \nn \\
   &=& 4 (m\tm)^2 \frac{\spb2.3 \spb1.4}{\spa2.3 \spa1.4} 
    =  4 (m\tm)^2 \frac{\spb1.2 \spb3.4}{\spa1.2 \spa3.4} \,.
\label{QuadCutAllPlusFinal}
\end{eqnarray}
For this helicity configuration, the parameter $t$ (which in the
quadruple cut takes on the value $t_0$ in \eqn{QuadCutt0}) drops out,
leading to a rather simple result.

To obtain dimensionally regularized results we need to adjust the
state sum.  Such adjustments have been discussed in
refs.~\cite{FDH,BernMorgan} and more recently in some detail in the
context of QCD calculations making use of six dimensions~\cite{GKM}.
Since we are working in six dimensions, there are $6-2 = 4$ gluon
states circulating in the loop.  This needs to be modified to match
dimensional regularization.  Two popular schemes are the
't~Hooft-Veltman scheme~\cite{HV}, where $2(1-\eps)$ gluon states
circulate in the loop, and the four-dimensional helicity (FDH)
scheme~\cite{FDH}, where two states circulate.  It is therefore
convenient to take the number of states to be $2 (1-\delta_R\eps)$,
where $\delta_R$ is unity for the 't~Hooft-Veltman scheme and zero for
the FDH scheme.

For our sample amplitude, since all four six-dimensional states give
identical contributions, the adjustment amounts to simply changing the
overall prefactor from $4\rightarrow 2 (1-\delta_R\eps)$. We also need
to adjust the number of components of extra-dimensional momenta from
six to $D=4 - 2 \eps$.  To do so we replace $m \tm \rightarrow \mu^2$,
where $\mu$ is understood to carry $(-2\eps)$-dimensional components
of momentum.  With these replacements, we obtain the
dimensionally regularized result for the quadruple cut,
\begin{equation}
C_{1234} = 2(1- \delta_R \eps) \mu^4 \frac{\spb1.2 \spb3.4}{\spa1.2 \spa3.4}\,.
\label{QuadCutAllPlusFinalDimReg}
\end{equation}

Since the identical helicity amplitude is determined entirely
from the quadruple cut, we obtain the dimensionally regularized
amplitude by putting back the four cut propagators and integrating
over the loop momentum in $4 -2\eps$ dimensions.  This gives,
\begin{equation}
A^\oneloop(1^+,2^+,3^+,4^+) = 2 (1 - \eps \delta_R) \frac{\spb1.2\spb3.4} 
  {\spa1.2 \spa3.4} I_4[\mu^4]\,,
\label{AllPlusAmplitude}
\end{equation}
where the box integral is
\begin{equation}
I_4[{\cal P}] = \int \frac{d^{4-2\eps} l}{(2 \pi)^{4-2\eps}}
  \frac{{\cal P}} {l^2 (l-p_1)^2 (l - p_1 -p_2)^2 (l +p_4)^2} \,,
\label{BoxIntegrals}
\end{equation}
with ${\cal P}$ a numerator polynomial.  This result
matches the known expression from ref.~\cite{BernMorgan}.

As a slightly more intricate example, consider the case where one
external gluon carries negative helicity and the others positive
helicity.  As before, the quadruple cuts and triple cuts are
conveniently extracted from the two-particle cuts, since these 
involve the simpler four-point tree amplitudes instead of three-point ones.
Following similar logic as for the identical-helicity case
(\ref{QuadCutAllPlusNaive}), for the present case
 the quadruple cut, illustrated in
\fig{OneLoopCutsFigure}(a), is given by,
\begin{eqnarray}
C_{1234} &=& 
 \frac{1} {s_{14} s_{23}} 
     \langle (-l_2)_a, 2_1, 3_1, (l_4)_b \rangle \,
            [ (-l_2)_\da, 2_{\dot{1}}, 3_{\dot{1}}, (l_4)_\db ]  \, \nn \\
 &&  \hskip 2 cm \times
           \langle (-l_4)^a, 4_1, 1_2, (l_2)^b \rangle \,
            [ (-l_4)^\da, 4_{\dot{1}}, 1_{\dot{2}}, (l_2)^\db ]   \,.
\end{eqnarray}
We then substitute in the solution for the spinor products, given in
\eqns{AllPlusSixD}{SingleMinusSixD}, appropriately relabeled and
with the parameter $t$ fixed to the quadruple cut value $t_0$ given in
\eqn{QuadCutt0}.  Summing over the internal helicity states then gives
\begin{eqnarray}
C_{1234}&=& -4 \frac{\kappa_{23}\kappa'_{23}\spa1.4^2\spb4.1^2
\spa3.2^2 \spb2.3^2\,t_0^2} {s_{23}^2} \nn \\ 
&=& 4 \frac{m \tm\spb2.4^2}{\spb1.2\spa2.3\spa3.4\spb4.1}
\frac{s_{12}s_{23}}{s_{24}}\left( \frac{s_{23}s_{12}}{2s_{24}}+m
\tm\pm \frac{s_{23}s_{12}}{2 s_{24}}\sqrt{1+ \frac{4 m \tm
s_{24}}{s_{23}s_{34}}}\right)\,.
\end{eqnarray}

As for the identical helicity case, each of the four states gives an
identical result, so the dimensionally regularized result is obtained
by replacing the overall prefactor of 4 by $2(1-\eps\delta_R)$ and
substituting $m \tm \rightarrow \mu^2$.  The box contribution is given
by the sum over both solutions (normalized by a factor of 1/2), so the
square root drops out.  Restoring the cut propagators and putting in
the loop integration gives the box contribution to the amplitude,
\begin{equation}
A(1^-, 2^+, 3^+, 4^+) \Bigr|_{\rm box} =  2 (1-\delta_R\eps)
\frac{\spb2.4^2}{\spb1.2\spa2.3\spa3.4\spb4.1}\frac{s_{12}
  s_{23}}{s_{13}} \left( \frac{s_{12}s_{23}}{2 s_{13}} I_4[\mu^2] +
  I_4[\mu^4] \right)\,,
\label{SingleMinusBoxes}
\end{equation}
where the box integrals are defined in \eqn{BoxIntegrals}.

The triple cut of \fig{OneLoopCutsFigure}(b) is also simple
to obtain from the two-particle cut in  \fig{OneLoopCutsFigure}(c) by
grouping the three-point trees together. After modifying
the state sum this gives,
\begin{eqnarray}
C_{124} &=& -2 (1-\eps\delta_R)
\frac{\kappa_{23}\kappa_{23}'\spa1.4^2\spb4.1^2 \spa2.3^2
\spb3.2^2\,t^2} {s_{23}^2} \frac{i}{(l_1 - p_1 - p_2)^2} \nn\\ &=& -2i
(1-\eps\delta_R) m \tm\spb2.3^2 t^2 \Bigl(t \, \spa4.2\spb1.2 + s_{12}
- \frac{m \tm}{t s_{14}} \spa1.2\spb4.2 \Bigr)^{-1} \,.
\end{eqnarray}
We used \eqn{TripleCutMomentum} to express $l_1 - p_1 - p_2$ in terms
of the parameter $t$.  In this form the corresponding triangle
coefficient can be extracted from the triple cut following
refs.~\cite{Forde,Badger}.  Similarly, starting from the two-particle
cuts, we can obtain the coefficients of bubble integrals.  With this
procedure, we can keep all terms which contribute to rational terms in
the amplitudes.

In summary, this illustrates the use of six-dimensional helicity in
unitarity cuts to obtain generalized cuts of one-loop amplitudes in
QCD, which can then be reduced to a basis set of integrals using
on-shell methods~\cite{OPP,GKM,Forde,Badger}.  Further details
for converting state sums from six dimensions to their
dimensionally regularized values may be found in ref.~\cite{GKM}.

\section{Tree-level maximally supersymmetric Yang-Mills amplitudes}
\label{SuperHelicitySection}

In this section we will describe tree-level amplitudes in $\NeqFour$
sYM theory using the recently constructed DHS on-shell superspace in
six dimensions~\cite{DHS}, as a precursor to carrying out multiloop
calculations via the unitarity method in the next section.  In
particular, we will describe a convenient means for setting up
high-point super-BCFW recursion for use in unitarity cuts.

\subsection{Tree-level amplitudes}

In the DHS superspace formalism, the on-shell superfield appears as a
polynomial in a pair of anticommuting coordinates
$\eta_a,\tilde{\eta}^{\dot{a}}$ which carry SU(2)$\times$SU(2) little
group indices,
\begin{eqnarray}
\Phi(\eta,\tilde{\eta}) &=&
  \phi 
  + \chi^a \eta_a 
  + \phi'(\eta)^2 
  + \tilde{\chi}_{\da}\tilde{\eta}^{\da} 
  + g^a\,_{\da}\eta_a\tilde{\eta}^{\da}
  + \tilde{\psi}_{\da}(\eta)^2\tilde{\eta}^{\da}\nonumber\\
  && \null
  + \phi''(\tilde{\eta})^2
  + \psi^a\eta_a(\tilde{\eta})^2
  + \phi'''(\eta)^2(\tilde{\eta})^2 \,,
\end{eqnarray}
where $(\eta)^2 \equiv\tfrac{1}{2}\epsilon^{ab}\eta_b\eta_a$ and
$(\tilde{\eta})^2 \equiv\tfrac{1}{2}\epsilon_{\dot{a}\dot{b}}
\tilde{\eta}^{\db}\tilde{\eta}^{\da}$.  Using these fields,
we can obtain superamplitudes in the usual way. 
 The different component
amplitudes can be read off from their $\eta$ expansions.  For example,
the four-gluon amplitude $\langle g^{a}{}_{\da}(1)\, g^{b}{}_{\db}(2)\,
g^{c}{}_{\dc}(3)\, g^{d}{}_{\dd}(4) \rangle$ appears as the coefficient
of $\eta_{1a}\tilde{\eta}_{1}^{\da} \eta_{2b}\tilde{\eta}_{2}^{\db}
\eta_{3c}\tilde{\eta}_{3}^{\dc} \eta_{4d}\tilde{\eta}_{4}^{\dd}$ in
the four-point superamplitude $\mathcal{A}_4$. These amplitudes are
functions of momenta $p_i$ and supermomenta $q_i,\tilde{q}_i$ defined
by,
\begin{equation}
  q_i^A= \lambda^{Aa}_i\eta_{ia},\qquad \tilde{q}_{iB}
       = \tilde{\lambda}_{iB\db}\tilde{\eta}_{i}^{\db} \,.
\end{equation}
For a general
   discussion of higher-dimensional on-shell superspaces, see
   ref.~\cite{Boels}.

In ref.~\cite{DHS}, the three-, four- and five-particle tree-level
superamplitudes were worked out analytically.
Consider first the four-point superamplitude, which is the 
simplest case. This is given by
\begin{eqnarray}
\mathcal{A}_{4}^{\text{tree}}(1,2,3,4)&=&
-\frac{i}{s_{12} s_{23}} \, \delta^4\biggl(\sum_{i=1}^4 q_i^A\biggr)
   \delta^4\biggl(\sum_{i=1}^4 \tilde{q}_{iB}\biggr)\,,
\label{FourPointSuperAmplitude}
\end{eqnarray}
where the fermionic delta function is defined as,
\begin{equation}
\delta^4 \Bigl({\nolsum_i} q_i^A\Bigr) \equiv
\frac{1}{4!} \, \e_{BCDE} \, \Bigl( {\nolsum_i} q_i^B\Bigr) 
\Bigl({\nolsum_i} q_i^C\Bigr) \Bigl({\nolsum_i} q_i^D\Bigr) 
\Bigl({\nolsum_i} q_i^E\Bigr) \,,
\end{equation}
and likewise for the
antichiral supermomentum $\tilde{q}_A$. 
This expression makes manifest the chiral--antichiral structure 
noted in \sect{DoubleCopySubsection} for the non-supersymmetric case.

The three-point superamplitude is a bit more complicated and is given 
by,
\begin{eqnarray}
\null \hskip -1.3 cm 
 \mathcal{A}_{3}^{\text{tree}}(1,2,3) &=&
-i \bigl(\bu_{1}\bu_{2}+
    \bu_{2}\bu_{3}+
    \bu_{3}\bu_{1}\bigr)
  \biggl(\sum_{i=1}^3 \bw_i\biggr)
  \bigl(\mathbf{\tilde{u}}_{1}\mathbf{\tilde{u}}_{2}+
    \btu_{2}\btu_{3}+
    \btu_{3}\btu_{1}\bigr)
    \biggl(\sum_{i=1}^3 \btw_i\biggr) \,,
\label{ThreePointSuperAmplitude}
\end{eqnarray}
where $\bu_i$ and $\bw_i$ are defined in terms of the $u_i^a$ and
$w_i^a$ of \app{ThreePointAppendix} as,
\begin{eqnarray}
 \bu_i = u_i^a\eta_{ia}, \quad 
    \btu_i = \tilde{u}_{i\da}\tilde{\eta}_{i}^{\da}, \quad 
    \bw_i=w_i^a\eta_{ia}, \quad
    \btw_i = \tilde{w}_{i\da}\tilde{\eta}_{i}^{\da} \,.
\end{eqnarray}
The expression in \eqn{ThreePointSuperAmplitude} is equivalent to the
form given in ref.~\cite{DHS}, except that here the chiral-conjugate 
property has been made manifest; it is a product of a chiral factor
and an antichiral factor. As discussed in \sect{DoubleCopySection}, 
this property survives the BCFW recursion, since the states
in the six-dimensional sYM theory factorize into a direct
product.  

The five-point tree-level superamplitude is %
\begin{eqnarray}
\nonumber\mathcal{A}_5^{\text{tree}}&=&\;
i \,\frac{\delta^4\Bigl(\sum_i
  q_i^A\Bigr) \delta^4\Bigl(\sum_i\tilde{q}_{iB} \Bigr)}
  {s_{12}s_{23}s_{34}s_{45}s_{51}}\Bigl\{q_{1}^A(p_2 p_3 p_4
  p_5)_A^{\;\;B}\tilde{q}_{1B}+\text{cyclic}\\ &&
  +\frac{1}{2}\left[q_{1}^A\tilde{\Delta}_{2A} +
  q_{3}^A\tilde{\Delta}_{4A} + (q_3+q_{4})^A\tilde{\Delta}_{5A} +
  (\text{chiral conjugate})\right]\Bigr\}\,,
\label{supertrees}
\end{eqnarray}
where $\tilde{\Delta}_{2A} = (p_2 p_3 p_4 p_5-p_2 p_5p_4
p_3)_{A}^{\;\;B}\tilde{q}_{2B}$, 
$\tilde{\Delta}_{4A} = (p_4 p_5 p_1 p_2-p_4 p_2 p_1
p_5)_{A}^{\;\;B}\tilde{q}_{4B}$, 
etc. The five-particle amplitude is
given here in a particularly compact form which lacks explicit cyclic
symmetry, although the symmetry does hold on the support of the
fermionic delta functions.

An important distinction between the DHS on-shell superspace and the
four-dimensional daughter theory is that this superspace is
non-chiral. The interesting consequence of this feature is that while
the four-dimensional amplitudes are organized according to N$^{k}$MHV
categories, these all come from a single six-dimensional parent
amplitude. This simplicity, however, is offset by the fact that one
lacks the counterpart of the four-dimensional MHV amplitudes, which
have a simple form for arbitrary numbers of legs.

\subsection{Supersymmetric BCFW recursion}
\label{SuperBCFWSubsection}

In order to investigate high loop orders we need high-point tree
amplitudes.  To obtain these, we use supersymmetric forms of the BCFW
recursion relations.  This also serves as a warmup for the unitarity
cuts, because the state sums generated by integration over Grassmann
parameters at loop level are similar to the state sums in 
supersymmetric forms of the BCFW recursion relations.
As discussed in ref.~\cite{DHS}, the super-BCFW
recursion relations can be obtained by shifting the Grassmann variables, 
\begin{eqnarray}
\nonumber\eta_{{1}a}(z)&=&\eta_{1a}-zX_{a\da}[1^{\da}|2^b\rangle\eta_{2b}/s_{12}\,, \hskip 2 cm 
\nonumber\eta_{{2}b}(z)=\eta_{2b}-zX^{a}_{\;\;\da}[1^{\da}|2_b\rangle
\eta_{1a}/s_{12} \,,\\
\tilde{\eta}_{{1}}^{\da}(z)&=&\tilde{\eta}_{1}^{\da}+zX^{a\da}[2_{\db}|1_a\rangle\tilde{\eta}_{2}^{\db}/s_{12}\,, \hskip 2.4cm 
\tilde{\eta}_{{2}}^{\db}(z)=\tilde{\eta}_{2}^{\db}+zX^{a}_{\;\;\da}[2^{\db}|1_a\rangle\tilde{\eta}_{1}^{\da}/s_{12} \,,
\label{fshift}
\end{eqnarray}
in addition to the shifts of the spinors in \eqn{bshift}.  This is
designed to maintain supermomentum conservation,
i.e. $q^A_1+q^A_2=q^A_{\hat{1}}+q^A_{\hat{2}}$ and
$\tilde{q}_{1A}+\tilde{q}_{2A}=\tilde{q}_{\hat{1}A}+\tilde{q}_{\hat{2}A}$.
These shifts on $\eta$ can be rephrased as shifts directly on the
supermomenta by combining with the spinorial shifts in \eqn{bshift}
and using the trick,
\begin{equation}
\delta^a_{\;\;b} = -\frac{1}{s_{12}}\langle 1^a | \s p_2 | 1_b \rangle \,.
\label{SpinorTrick}
\end{equation}
After the rearrangements, we have,
\begin{eqnarray}
q^A_{{1}}(z) &=& 
    q^A_{1} 
    + \frac{z}{s_{12}^2} X^{a}_{\;\;\da} [1^{\da}|2^b\rangle \lambda_{2b}^A \langle 1_a|2_{\db}] \tilde{\lambda}^{\db}_{2B} q^{B}_{1}
  - \frac{z}{s_{12}} \lambda_{1}^{Aa} X_{a\da}\tilde{\lambda}_{1B}^{\da}q_{2}^B\,,
              \nonumber\\
  q^A_{{2}}(z) &=& 
    q^A_{2}
    - \frac{z}{s_{12}} \lambda_{1a}^{A} X^{a}_{\;\;\da} \tilde{\lambda}_{1B}^{\da} q_2^B 
    - \frac{z}{s_{12}^2} X^{a}_{\;\;\da} [1^{\da}|2_b \rangle \lambda_{2}^{Ab} \langle 1_{a}| 2_{\db}] \tilde{\lambda}_{2B}^{\db} q_{1}^{B} \,, \nonumber \\
  \tilde{q}_{{1}A}(z) &=&
    \tilde{q}_{1A}
    + \frac{z}{s_{12}^2} X^{a}_{\;\;\da} {\langle 1_a|2_{\db}]} \tilde{\lambda}_{2A}^{\db} {[1^{\da}|2^b\rangle} \lambda_{2b}^B \tilde{q}_{1B}
    + \frac{z}{s_{12}} \tilde{\lambda}_{1A\da} X^{a\da} \lambda_{1a}^{B} \tilde{q}_{2B} \,, \nonumber \\
  \tilde{q}_{{2}A} (z) &=&
    \tilde{q}_{2A}
    + \frac{z}{s_{12}} X^{a}_{\;\;\da} \tilde{\lambda}_{1A}^{\da} \lambda_{1a}^{B} \tilde{q}_{2B}
    + \frac{z}{s_{12}^2} X^{a}_{\;\;\da} \tilde{\lambda}_{2A\db}[2^{\db}|1_a\rangle [1^\da|2^b\rangle \lambda_{2b}^{B} \tilde{q}_{1B} \,.
\end{eqnarray}
As we will see, these supermomentum shifts help us avoid dealing 
directly with the supercoordinates $\eta$ in the recursion.

The intermediate state sum in the recursion is realized as an
integration over the Grassmann coordinates $\eta_P,\tilde{\eta}_P$ of
the intermediate leg (labeled as $P$), and the remainder of this
section is devoted to systematically carrying out these integrations.

To set up high-point recursion, it is useful to first organize
the types of contributions.
If we track the factors containing Grassmann parameters, and drop 
other factors,  the $n$-point tree amplitudes have the schematic form,
\begin{eqnarray}
\mathcal{A}_n&\sim &\delta^4\biggl({\nolsum_i} q_i^A\biggr)
  \delta^4\Bigl({\nolsum_i}\tilde{q}_{iB}\Bigr)
  q^{n-4}\tilde{q}^{n-4}\quad{\rm for}\; n\ge 4 \,.
\label{face}
\end{eqnarray}
The supermomentum delta functions for $n\ge4$ impose algebraic
constraints on $\eta_P,\tilde{\eta}_P$ under the fermionic
integration.  We can follow the same strategy as used in four
dimensions to consider the delta-function constraints as a set of
algebraic equations to be systematically solved~\cite{SuperSum}.
A key identity for carrying this out is
\begin{eqnarray}
\delta^{4}(q_1^A+q_2^A+Q^A) &=& s_{12} \,
  \delta^2\Bigl(\eta_{1a}+s_{12}^{-1}\langle1_a|\s{p}_2Q\Bigr) 	
  \delta^2\Bigl(\eta_{2b}+s_{12}^{-1}\langle 2_b|\s{p}_1Q\Bigr)\,.
\label{deltaidentity}
\end{eqnarray}
A similar antichiral identity also holds. These identities are
analogous to ones used in four dimensions~\cite{DeltaIdentity}.

\begin{figure}
\begin{center}
\centerline{\epsfxsize 4.5truein \epsfbox{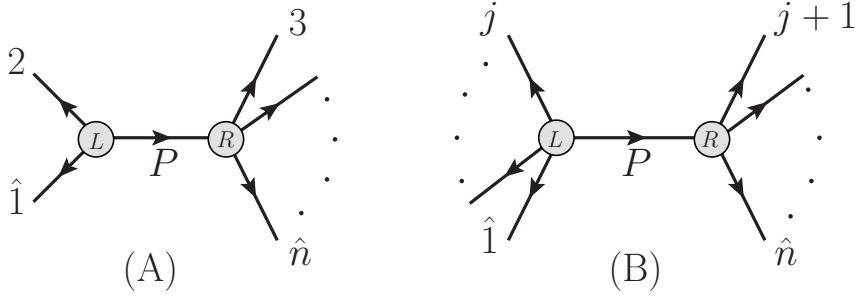}}
\caption{Two categories of BCFW diagrams. (A) contains a three-point
subamplitude, which does not have the full supermomentum delta
function $\delta^4(q)\delta^4(\tilde{q})$. (B) contains no three-point
subamplitude, thus all delta functions are of degree four.
}
\label{BCFWcategories}
\end{center}
\end{figure}

Before proceeding, though, we are obliged to address a glitch in
the spinor-helicity formalism: the spinors do not properly distinguish
between particles and antiparticles, causing phase inconsistencies in
diagrams containing fermions.  This glitch and prescriptions for
resolving it have already been discussed in the four-dimensional case,
for example, in ref.~\cite{FermionGlitch,SuperSum}. Here we use the
prescription that if $p$ is incoming, so that this momentum is $-p$ in
our all-outgoing convention, then we take,
\begin{equation}
 \lambda^{Aa}_{(-p)}\equiv i\lambda^{Aa}_{p},\qquad 
\tilde{\lambda}_{(-p)A\da} \equiv i\tilde{\lambda}_{p A\da} \,.
\end{equation}
In this way, we maintain the relationship between momenta and spinors, i.e.,
\begin{eqnarray}
 (-p)^{AB} = \lambda_{(-p)}^{Aa}\lambda_{(-p)a}^B 
 = -\lambda_{p}^{Aa}\lambda_{pa}^B \,.
\end{eqnarray}
Along the same lines, we demand that $q_{(-p)}^A = -q_{p}^A$, which we
achieve with the same prescription for the supercoordinates $\eta$:
whenever $-p$ is outgoing, we choose,
\begin{align}
 \eta_{(-p)a}\equiv i\eta_{pa},\qquad 
 \tilde{\eta}_{(-p)}^{\da} \equiv i\tilde{\eta}_{p}^{\da}.
\end{align}

Beyond four points, we can split all BCFW diagrams into two categories,
as shown in \fig{BCFWcategories}.  We consider these two cases in
turn.
\vskip .3 cm 
\noindent
\underline{\bf Case (A):}

\vskip .2 cm 

In cases where one has to sew a three-point superamplitude to a
higher-point tree superamplitude, illustrated in \fig{BCFWcategories}(A),
the Grassmann integration has the form,
\begin{eqnarray}
&&  \int d^2\eta_Pd^2\tilde{\eta}_P\; \mathcal{A}_{3L} \times
  \delta^4\biggl(\sum_{i \in R} \! q_i^B\biggr)\delta^4
   \biggl(\sum_{i \in R} \! \tilde{q}_{iC}\biggr)q^{n-5}\tilde{q}^{n-5} \,,
  \end{eqnarray}
where the summations are over all external lines of the right tree
amplitude including shifted legs, and the integration measure is
$d^2\eta_P d^2\tilde{\eta}_P =
\tfrac{1}{4}d\eta_{P}^{a}d\eta_{Pa}d\tilde{\eta}_{P\da}d\tilde{\eta}_{P}^{\da}$.

To perform this integral, we view the factors in the three-point
amplitude (\ref{ThreePointSuperAmplitude}) as algebraic constraints on
the supercoordinates $\eta_P,\tilde{\eta}_P$ imposing,
\begin{eqnarray}
  &&\bu_{\hat{1}} = \bu_2 \,,\qquad \bu_P 
= \tfrac{1}{2}(\bu_{\hat{1}}+\bu_2),\qquad \bw_P=-\bw_{\hat{1}}-\bw_2, \nonumber\\
  &&\btu_{\hat{1}} = \btu_2 \,,\qquad \btu_P 
= \tfrac{1}{2}(\btu_{\hat{1}}+\btu_2),\qquad \btw_P=-\btw_{\hat{1}}-\btw_2\,.
\end{eqnarray}
Since there are two components each in $\eta_{Pa}$ and 
$\tilde{\eta}_{P}^{\da}$, the three-point amplitude is sufficient to
localize the $\eta_P$ integrals. The solutions to the constraint equations are
\begin{eqnarray}
 \eta_{Pa} &=& \tfrac{1}{2}w_{Pa}(\bu_{\hat{1}}+\bu_2) 
    + u_{Pa}(\bw_{\hat{1}}+\bw_2) \,, \nonumber\\
  \tilde{\eta}_{P}^{\da} 
&=& -\tfrac{1}{2}\tilde{w}_{P}^{\da}(\btu_{\hat{1}}+\btu_2)
    - \tilde{u}_{P}^{\da}(\btw_{\hat{1}}+\btw_2) \,,
\label{ConstraintSolutions}
\end{eqnarray}
with the understanding that we are free to replace
 $\bu_{\hat{1}}\leftrightarrow\bu_2$ and $\btu_{\hat{1}}\leftrightarrow\btu_2$
at any step.

In practice, it is much easier to deal with supermomenta $q_i$ than
the $u$ and $w$ variables.  It is straightforward to demonstrate
that \eqn{ConstraintSolutions} implies the substitutions,
\begin{eqnarray}
 q^A_P = -q^A_{\hat{1}}-q^A_2\,, \hskip 1.5 cm 
\tilde{q}_{PA} = -\tilde{q}_{\hat{1}A}-\tilde{q}_{2A} \,.
\label{qSubstitution}
\end{eqnarray}
Thus one can substitute the result for
$q_P$ in the tree amplitude on the right, avoiding the more complicated \eqn{ConstraintSolutions}, and extract out the full
supermomentum-conservation delta function.
At this stage, we are left with the task of integrating,
\begin{eqnarray}
    &&\delta^4\biggl(\sum_{i \in \E } q^A_i\biggr)
   \delta^4\biggl(\sum_{i \in \E}\tilde{q}_{iB}\biggr)
        q^{n-5}\tilde{q}^{n-5} \int d^2\eta_Pd^2\tilde{\eta}_P
    \mathcal{A}_{3L} \,,
\end{eqnarray}
where the two delta functions correspond to the overall supermomentum
conservation, and $\E$ is the set of all external legs of the full
amplitude. The remaining integral has the solution,
\begin{equation}
  \int d^2\eta_Pd^2\tilde{\eta}_P\mathcal{A}_{3L}=i(\bu_{\hat{1}}-\bu_{2})(\btu_{\hat{1}}-\btu_{2}) \,,
\end{equation}
which can be re-written in terms of the $q,\tilde{q}$ as,
\begin{equation}
i\biggl(
-\frac{q_{\hat{1}}\s{p}_K\s{p}_2\tilde{q}_{\hat{1}}}{s_{\hat{1}K}}
-q_{\hat{1}}\tilde{q}_2
+q_2\tilde{q}_{\hat{1}}
+\frac{q_2\s{p}_K\s{p}_{\hat{1}}\tilde{q}_2}{s_{2K}}
\biggr) \,,
\end{equation}
where $p_K$ is an arbitrary null reference vector.
The sewing of a three-point amplitude with a general tree amplitude
will thus result in the form,
\begin{equation}
\delta^4\biggl(\sum_{i\in \E} q_i ^A\biggr)
\delta^4\biggl(\sum_{i\in \E}\tilde{q}_{iB}\biggr)q^{n-5}\tilde{q}^{n-5}
  \biggl(
    -\frac{q_{\hat{1}}\s{p}_K\s{p}_2\tilde{q}_{\hat{1}}}{s_{\hat{1}K}}
    -q_{\hat{1}}\tilde{q}_2
    +q_2\tilde{q}_{\hat{1}}
    +\frac{q_2\s{p}_K\s{p}_{\hat{1}}\tilde{q}_2}{s_{2K}}
  \biggr)\,,
\end{equation}
where all $q,\tilde{q}$s are in terms of external lines and  $q_{\hat{1}}$
is the shifted $q_1$ with $z$ taking on the value at the pole.

\vskip .3 cm 
\noindent
\underline{\bf Case (B):}

\vskip .2 cm 

For the case with no three-point subamplitudes, illustrated in
\fig{BCFWcategories}(B), we can always eliminate  $q_P$ and $\tilde{q}_P$
from the ``$R$'' amplitude, using the ``$L$'' delta functions.
The Grassmann integral will then be of the form,
\begin{equation}
\delta^4\biggl(\sum_{i \in \E} q_i^A\biggr)
\delta^4\biggl(\sum_{i \in \E}\tilde{q}_{iB}\biggr)f(q,\tilde{q})
\int d^2\eta_Pd^2\tilde{\eta}_P \,
\delta^4 \biggl(\sum_{i \in L} q_i^C \biggr)
  \delta^4\biggl(\sum_{i \in L} \tilde{q}_{iD}\biggr)\,.
  \end{equation}
Focusing on the chiral integral, we proceed by splitting the delta
function using \eqn{deltaidentity} on legs $j$ and $P$, which 
gives,
\begin{eqnarray}
 \int d^2\eta_P\delta^4\biggl(\sum_{i \in L} q_i^C \biggr) 
  &=& s_{j P} \,\delta^2 \biggl(\eta_{ja} + s_{j P}^{-1}\langle j_a|{\s P}
 (q_{\hat{1}}+\ldots+q_{j-1})\biggr) \nonumber\\
  &=& s_{j P} \,\delta^2 \biggl(s_{j P}^{-1}\langle j_a| {\s P} (q_{\hat{1}}+\ldots+q_{j})\biggr) \nonumber\\
  &=& -s_{j P}^{-1} (q_{\hat{1}}+\ldots+q_{j}) {\s P} \s p_j {\s P} (q_{\hat{1}}+\ldots+q_{j}) \nonumber\\
  &=& (q_{\hat{1}}+\ldots+q_{j})^A (p_{\hat{1}}+\ldots+p_{j})_{AB} (q_{\hat{1}}+\ldots+q_{j})^B\,,
\end{eqnarray}
where $P= p_{\hat 1} + p_2 +\cdots + p_j$, and 
we used the identity (\ref{SpinorTrick}) between the first and
second lines.   Likewise, the antichiral integration contributes,
\begin{eqnarray}
 \int d^2\tilde{\eta}_P\delta^4\biggl(\sum_{i \in L} \tilde{q}_{iD}\biggr) 
  &=& (\tilde{q}_{\hat{1}}+\ldots+\tilde{q}_{j})_A (p_{\hat{1}}+\ldots+p_{j})^{AB}(\tilde{q}_{\hat{1}}+\ldots+\tilde{q}_{j})_B\,.
\end{eqnarray}
%

\section{Multiloop applications with maximal supersymmetry}
\label{MultiLoopApplicationSection}

\begin{figure}
\begin{center}
\centerline{\epsfxsize 3.truein \epsfbox{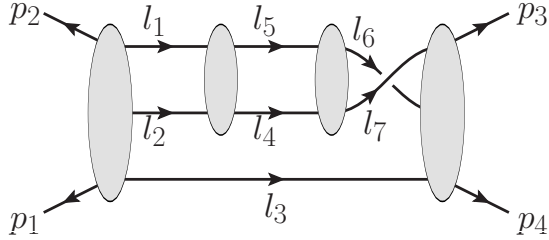}}
\caption{A sample cut of the four-loop four-point amplitude.}
\label{4LoopCut}
\end{center}
\end{figure}

In this section we consider the construction of 
multiloop amplitudes in maximally supersymmetric 
Yang-Mills in six dimensions. We consider the $(1,1)$ theory,
which, under a dimensional reduction to four dimensions, 
is equivalent to $\NeqFour$ sYM theory.  After
some general comments, we turn to a two-loop warmup
before discussing four-point amplitudes at four and six
loops. 

\subsection{General considerations}

In unitarity cuts, one must sum over states in multiple
lines.  The supersum then corresponds to integrating over the
$\eta,\tilde{\eta}$ coordinates of these lines. For example, when
sewing the four-loop cut displayed in \fig{4LoopCut},
the sum over states is implemented by the Grassmann integration,
\begin{equation} 
\int \prod_{i=1}^7
\biggl(d^2\eta_{l_i}d^2\tilde{\eta}_{l_i}\biggr)
   \mathcal{A}_5^{(1)}\mathcal{A}_4^{(2)}
     \mathcal{A}_4^{(3)}\mathcal{A}_5^{(4)} \,,
\end{equation}
where the superscript $\mathcal{A}_4^{(j)}$ labels the four distinct 
tree amplitudes composing the cut.  

The strategy for dealing with such supersums is similar to the
strategy used at tree level; we use \eqn{deltaidentity} to localize as
many $\eta$ integrals as possible.  In cuts with no three-point
subamplitudes, such as the cut in \fig{4LoopCut}, there is a
supermomentum delta function on each tree amplitude. Each delta
function can be used to localize two pairs of
$\eta_{l},\tilde{\eta}_l$ via \eqn{deltaidentity}; in a cut with $m$
tree amplitudes, a total of $2(m-1)$ pair of $\eta_{l},\tilde{\eta}_l$
can be localized in this manner, with one overall $\delta^4(\sum_{\E}
q)\delta^4(\sum_{\E}\tilde{q})$ extracted outside of the integral. We
note that when solving the delta-function constraints, care must be
taken to avoid circular solutions.

In general, the fermionic delta functions will be insufficient to
localize all of the $\eta$ integrals; the remaining integrals must be
handled in a different manner. One approach is to expand the left-over
integrand as a polynomial in $\eta_{l_i},\tilde{\eta}_{l_i}$ and
interpret the fermionic integral as instructions to pick out the
coefficient of $\prod_i(\eta_{l_i})^2(\tilde{\eta}_{l_i})^2$. We will
apply this approach to a two-loop example in the next subsection.

Cuts that have three-point subamplitudes are generally more
difficult, and may need to be handled on an \textit{ad hoc} basis.
One can usually make progress by combining three-point
subamplitudes into higher-point amplitudes, as was done in the
non-supersymmetric case in \sect{OneLoopQCDSection}. 
One can perform the remaining $\eta$
integrals as before.

We note that for four-point loop amplitudes, after extracting out the
overall supermomentum delta functions, the number of remaining
$\eta,\tilde{\eta}$s will always match the number of Grassmann
integrations. Therefore, the four-point loop amplitudes
will depend on $\eta,\tilde{\eta}$ only through the supermomentum
delta functions, i.e. they will be proportional to a four-point tree
superamplitude.

To illustrate these techniques we now work out a few examples.  (The
cut of \fig{4LoopCut} is evaluated in some detail in
\app{FourLoopCutK}.)

\subsection{Two-loop four-point example}
\label{TwoLoopFourPoint}

\begin{figure}
\begin{center}
\centerline{\epsfxsize 2.2 truein \epsfbox{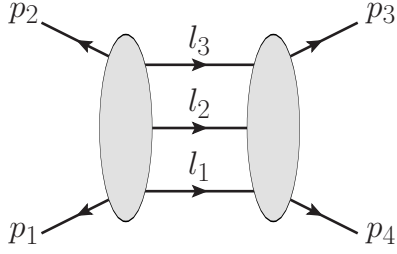}}
\caption{The three-particle cut of a two-loop four-point amplitude.}
\label{2loop3part}
\end{center}
\end{figure} 

We first evaluate the three-particle cut illustrated in
\fig{2loop3part}. The cut is given by sewing together two five-point
tree superamplitudes,
\begin{equation}
C^\twoloop = 
\int \prod_{i=1}^3 
d^2\eta_{l_i}d^2\tilde{\eta}_{l_i}
\, \mathcal{A}^\tree_{5L}(p_1,p_2,l_3,l_2,l_1)
\mathcal{A}^\tree_{5R}(p_3,p_4,-l_1,-l_2,-l_3) \,.
\label{2loopint}
\end{equation}
We choose the following convenient representations for the
five-point tree superamplitudes:
\begin{eqnarray}
\nonumber\mathcal{A}^\tree_{5L}&=& i\,\frac{\delta^4(\sum_L q)\delta^4(\sum_L\tilde{q})}{s_{l_3l_2}s_{l_2l_1}s_{l_11}s_{12}s_{2l_3}}\biggl\{q_{l_3}^A(l_2 l_1 p_1 p_2)_A^{\;\;B}\tilde{q}_{l_3B}+\text{cyclic}\\
\nonumber&& \null
+\frac{1}{2}\left[q_{l_3}^A\tilde{\Delta}^L_{l_2A} +q_{l_1}^A\tilde{\Delta}^L_{1A}+(q_{l_1}+q_{1})^A\tilde{\Delta}^L_{2A}+(\text{chiral conjugate})\right]\biggr\} \,,\\
\nonumber\mathcal{A}^\tree_{5R}&=& i\,\frac{\delta^4(\sum_R q)\delta^4(\sum_R\tilde{q})}{s_{l_1l_2}s_{l_2l_3}s_{l_33}s_{34}s_{4l_1}}\biggl\{q_{l_1}^A(l_2 l_3 p_3 p_4)_A^{\;\;B}\tilde{q}_{l_1B}+\text{cyclic}\\
&& \null
+\frac{1}{2}\left[-q_{l_1}^A\tilde{\Delta}^R_{(-l_2)A} -q_{l_3}^A\tilde{\Delta}^R_{3A}+(q_3-q_{l_3})^A\tilde{\Delta}^R_{4A}+(\text{chiral conjugate})\right]\biggr\} \,,
\label{rep}
\end{eqnarray}
where the $L,R$ superscript of the $\Delta$'s denote with respect to
which side of the cut they are defined. For example, we have,
\begin{eqnarray}
\nonumber\tilde{\Delta}^L_{l_2A} &=&(l_2 l_1 p_1 p_2-l_2p_2p_1l_1)_{A}^{\;\;B}\tilde{q}_{l_2B}\,, \\
\tilde{\Delta}^R_{(-l_2)A} &=&-(l_2 l_3 p_3 p_4-l_2p_4p_3l_3)_{A}^{\;\;B}\tilde{q}_{l_2B}\,.
\end{eqnarray}

As discussed previously, one can extract an overall supermomentum
delta function $\delta^4(\sum_{\E} q)\delta^4(\sum_{\E}\tilde{q})$
outside of the integral, leaving behind a degree eight delta function,
which can be used to localize two pairs of $\eta$s and $\tilde
\eta$s. Here we choose to localize $\eta_{l_1}$, $\eta_{l_2}$ and
their antichiral partners. Thus the fermionic delta functions from
the two tree amplitudes combine to give,
\begin{eqnarray}
\nonumber &&
\hskip -1.7 cm 
\delta^4\Bigl(\sum_L q\Bigr)
\delta^4\Bigl(\sum_L \tilde{q}\Bigr) 
\delta^4\Bigl(\sum_R q\Bigr)
\delta^4\Bigl(\sum_R \tilde{q}\Bigr)\\
\nonumber&=&\delta^4\Bigl(\sum_{\E} q\Bigr)
\delta^4\Bigl(\sum_{\E}\tilde{q}\Bigr)s^2_{l_1l_2} \\
\nonumber && \times
\delta^2\Bigl(\eta_{l_1a}+s_{l_1l_2}^{-1}\langle l_{1a}|\s{l}_{2}(q_{l_3}+q_1+q_2)\Bigr)
\delta^2\Bigl(\eta_{l_2b}+s_{l_1l_2}^{-1}\langle l_{2b}|\s{l}_{1}(q_{l_3}+q_1+q_2)\Bigr)\\
&& \times
\delta^2\Bigl(\tilde{\eta}_{l_1}^{\da}+s_{l_1l_2}^{-1} [l_{1}^{\da}|\s{l}_{2}(\tilde{q}_{l_3}+\tilde{q}_1+\tilde{q}_2)\Bigr)
\delta^2\Bigl(\tilde{\eta}_{l_2}^{\db}+s_{l_1l_2}^{-1}[l_{2}^{\db}|\s{l}_{1}(\tilde{q}_{l_3}+\tilde{q}_1+\tilde{q}_2)\Bigr) \,,
\label{local}
\end{eqnarray}
which is a direct application of \eqn{deltaidentity}. From the above,
one can immediately see that the delta function localizes the
$d^2\eta_{l_i}d^2\tilde{\eta}_{l_i}$ integral in \eqn{2loopint} for
$i=1,2$.

After factoring out the fermionic delta functions in
$\mathcal{A}_{5L}\mathcal{A}_{5R}$, the remaining function is of
Grassmann degree four, and only terms proportional to
$(\eta_{l_3})^2 (\tilde{\eta}_{l_3})^2$ can saturate the
remaining Grassmann integrals. Keeping in mind that
$\eta_{l_1},\eta_{l_2},\tilde{\eta}_{l_1},\tilde{\eta}_{l_2}$ are
localized through \eqn{local}, the contributing terms in the curly
bracket in \eqn{rep} are,
\begin{eqnarray}
\nonumber&&\hskip -.3 cm
\biggl\{q_{l_3}^A(l_2 l_1 p_1 p_2)_A^{\;\;B}\tilde{q}_{l_3B}+q_{l_2}^A(l_1 p_1
p_2l_3)_A^{\;\;B}\tilde{q}_{l_2B}+q_{l_1}^A( p_1 p_2l_3l_2)_A^{\;\;B}\tilde{q}_{l_1B}
+\frac{1}{2}\left[q_{l_3}^A\tilde{\Delta}^L_{l_2A}
+\tilde{q}_{l_3A}\Delta^{LA}_{l_2}\right]\biggr\}\\
\nonumber&& \hskip -.6 cm\times
\biggl\{q_{l_1}^A(l_2 l_3 p_3 p_4)_A^{\;\;B}\tilde{q}_{l_1B}+q_{l_2}^A( l_3 p_3
p_4l_1)_A^{\;\;B}\tilde{q}_{l_2B}
+q_{l_3}^A(p_3p_4l_1l_2)_A^{\;\;B}\tilde{q}_{l_3B}-\frac{1}{2}\left[q_{l_1}^A\tilde{\Delta}^R_{(-l_2)A}+\tilde{q}_{l_A}\Delta^{RA}_{(-l_2)}\right]\biggr\}\,.\\
\end{eqnarray}
Thus performing the final $d^2\eta_{l_3}d^2\tilde{\eta}_{l_3}$ integration
 gives for \eqn{2loopint},
\begin{eqnarray}
\nonumber&& \hskip -2. cm 
C^\twoloop =
\frac{i s_{23}\mathcal{A}^{\tree}_4(p_1,p_2,p_3,p_4)}
     {s_{12} (l_3+l_2)^4 (l_1+p_1)^2 (p_2+l_3)^2 (l_3-p_3)^2 (p_4-l_1)^2} \\
\nonumber&\times& \hskip -.2 cm 
\biggl\{\langle l_3 | \s{l}_2 \s{l}_1 \s{p}_1 \s{p}_2+\frac{\s{l}_1 \s{p}_1 \s{p}_2\s{l}_3\s{l}_2\s{l}_1}{(l_1+l_2)^2}+\frac{\s{l}_2\s{l}_1 \s{p}_1 \s{p}_2\s{l}_3\s{l}_2}{(l_1+l_2)^2} | l_3]\\
\nonumber&&-\frac{1}{2}\left[\langle l_3|\frac{(\s{l}_2 \s{l}_1 \s{p}_1 \s{p}_2-\s{l}_2\s{p}_2\s{p}_1\s{l}_1)\s{l}_2\s{l}_1}{(l_1+l_2)^2}|l_3]-
  [l_3|\frac{(\s{l}_2 \s{l}_1 \s{p}_1 \s{p}_2-\s{l}_2\s{p}_2\s{p}_1\s{l}_1)\s{l}_2\s{l}_1}{(l_1+l_2)^2}|l_3\rangle\right]\biggr\}^{a}_{\;\;\dot{a}}\\
\nonumber&\times&  \hskip -.2 cm 
\biggl\{\langle l_3|\frac{\s{l}_2 \s{l}_3 \s{p}_3 \s{p}_4\s{l}_1\s{l}_2}{(l_1+l_2)^2}+
  \frac{\s{l}_1\s{l}_2 \s{l}_3 \s{p}_3 \s{p}_4\s{l}_1}{(l_1+l_2)^2}+
  \s{p}_3 \s{p}_4\s{l}_1\s{l}_2|l_3]\\
&&+\frac{1}{2}\left[\langle l_3|\frac{(\s{l}_2 \s{l}_3 \s{p}_3 \s{p}_4-\s{l}_2\s{p}_4\s{p}_3\s{l}_3)\s{l}_2\s{l}_1}{(l_1+l_2)^2}|l_3]-
  [l_3|\frac{(\s{l}_2 \s{l}_3 \s{p}_3 \s{p}_4-\s{l}_2\s{p}_4\s{p}_3\s{l}_3)\s{l}_2\s{l}_1}{(l_1+l_2)^2}|l_3\rangle\right]\biggr\}_{a}^{\;\;\dot{a}} \,.
\label{TwoLoopCutResult}
\end{eqnarray}
This expression can be cleaned up further, but for our purposes it is
simplest to numerically evaluate it.  We have numerically checked that
after dividing by the tree amplitude, this expression matches the
analogous expression obtained using four-dimensional
cuts~\cite{BRY,BDDPR}, but extended into six dimensions.  Since the
four-dimensional expression depends only on Lorentz dot products of
momenta, this extension is carried out simply by treating the dot
products as six-dimensional ones.

\subsection{Multiloop $\NeqFour$ super-Yang-Mills and $\NeqEight$ supergravity}
\label{FourloopFourpointSubsection}

Following the procedure described above, one can directly check the
six-dimensional unitarity cuts of more complicated multiloop
$\NeqFour$ sYM amplitudes.  Up to three loops, the four-gluon amplitude
is known to be valid in $D$ dimensions (subject to mild power counting
assumptions)~\cite{GravityThree,CompactThree}.
As a nontrivial application of the methods described above, here
we confirm that the complete four-loop four-particle amplitudes of
$\NeqFour$ sYM theory  computed in ref.~\cite{FourLoop} are indeed
valid for $D \le 6$.  In that paper, the amplitude was given as a
linear combination of 50 integrals of the form,
\begin{equation}
st A_4^\tree \int \left(\prod_{i = 1}^4 \frac{d^D l_i}{(2\pi)^D}\right)
            \frac{N_k(l_i, p_i)} { \prod_{j = 1}^{16} l_j^2} \,,
\label{FourLoopForm}
\end{equation}
where the numerator $N_k$ is a polynomial of degree six in the loop and external
momenta. Of these integrals, six are planar and
the rest nonplanar.  In ref.~\cite{FourLoop}, many of the terms in the
numerators were explicitly determined using cuts with $D=4$
momenta and helicity states.  Although a number of nontrivial checks
were performed partially confirming their validity in $D>4$ dimensions,
it is still useful to have a complete confirmation valid especially
for $D=11/2$, which is the lowest dimension where an ultraviolet
divergence can occur. 

As explained in ref.~\cite{GravityFour}, the validity of $\NeqFour$
sYM amplitudes in $D$ dimensions implies that the
corresponding $\NeqEight$ supergravity amplitudes are valid as well.
This follows from the construction of $\NeqEight$ supergravity
amplitudes from corresponding $\NeqFour$ sYM amplitudes using the
unitarity method in conjunction with the KLT relations~\cite{KLT},
which are known to be valid in $D$ dimensions.  We therefore need only
confirm the $D=6$ validity of the $\Neqfour$ sYM amplitudes to confirm
the result for $\NeqEight$ supergravity for $4 < D \le 6$.  A key conclusion
of ref.~\cite{GravityFour} is that the four-loop four-point
amplitude of $\NeqEight$ supergravity then cannot diverge in dimensions lower
than  $D=11/2$, matching the behavior of $\NeqFour$ sYM theory.

\begin{figure}
\begin{center}
\centerline{\epsfxsize 4.5truein \epsfbox{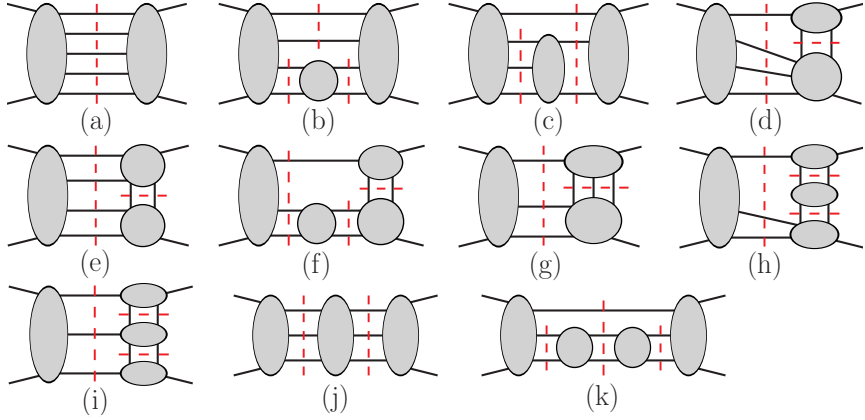}}
\caption{The eleven basic cuts decomposing a four-point four-loop
  amplitude into a product of tree amplitudes.  Together with ones
  involving two-particle cuts, these are the basic ones for determining
  a massless four-point four-loop amplitude, including nonplanar
  contributions~\cite{GravityFour,FourLoop}.  The complete spanning set is
  given by taking all possible distinct permutations of cut and
  external legs.}
\label{CutBasisFigure}
\end{center}
\end{figure}

As discussed in refs.~\cite{GravityFour,FourLoop}, any four-point
four-loop amplitude of a massless theory can be completely determined
(up to scale-free integrals that integrate to zero in dimensional
regularization) via a set of eleven basic cuts shown in
\fig{CutBasisFigure} along with simpler ones containing two-particle
cuts, not displayed. The complete spanning set is obtained by
considering all permutations of the legs of each constituent tree
amplitude.  This set is constructed by demanding that all potential
terms are detectable in at least one cut.  To carry out our
six-dimensional evaluation, we compare the cuts formed from the
products of tree amplitudes against the corresponding cuts of the
amplitudes as given in ref.~\cite{FourLoop}.  In the latter form,
after dividing by the tree amplitude, only Lorentz inner products
remain in the amplitude, making
numerical comparisons in six dimensions straightforward.  In
\app{FourLoopCutK}, we give an explicit analytic evaluation of the
sample nonplanar cut shown in \fig{4LoopCut}.

More generally, after extracting an overall supermomentum-conservation
delta function from the cuts, the remaining delta functions localize
most of the $\eta\tilde{\eta}$ integration, leaving behind six, four,
and two pairs of $\eta\tilde{\eta}$ respectively for diagrams (a),
(b,c,d,e,g,j), and (f,h,i,k). These extra pairs of $\eta\tilde{\eta}$s
then saturate the remaining integration, as was the case in the
two-loop example above.  This then gives us a result for the state sum
of these cuts, which we evaluate with six-dimensional kinematics
constrained to satisfy the on-shell conditions.

We find appropriate numerical solutions to the cut conditions by
sequentially building momenta that satisfy the on-shell conditions of
each of the constituent tree amplitudes in a cut.  For a constituent
$n$-point tree amplitude, we need to impose momentum conservation and take
the momenta on shell, $p_i^2=0$.  For $n-2$ of the legs, we can choose
arbitrary null vectors, imposing the momentum conservation constraints
on the final two legs. We label the sum of the arbitrary $n-2$ momenta
as $P=\sum_i^{n-2}p_i$.  We then define $p_{n-1}\equiv \left(-{P^2}/{2
  k \cdot P}\right) k $, where $k$ is an arbitrary null vector.  With
this choice then $p_{n} \equiv p_{n-1}+K$ automatically satisfies the
on-shell condition $p_{n}^2 =0$.  (It can happen that momentum
conservation constraints can lead to inconsistencies in this simple
procedure, if a tree does not have at least two unspecified legs.
Although this can always be avoided in our case, we note that such
inconsistencies can be resolved by solving the cut conditions for
groups of tree amplitudes instead of one by one.)

By systematically stepping through the spanning set of cuts described
above, we confirm that for $D\le 6$ the full amplitude is correctly
constructed by the mostly four-dimensional evaluation of
ref.~\cite{FourLoop}, as expected.

\begin{figure}
\begin{center}
\centerline{\epsfxsize 2.5 truein \epsfbox{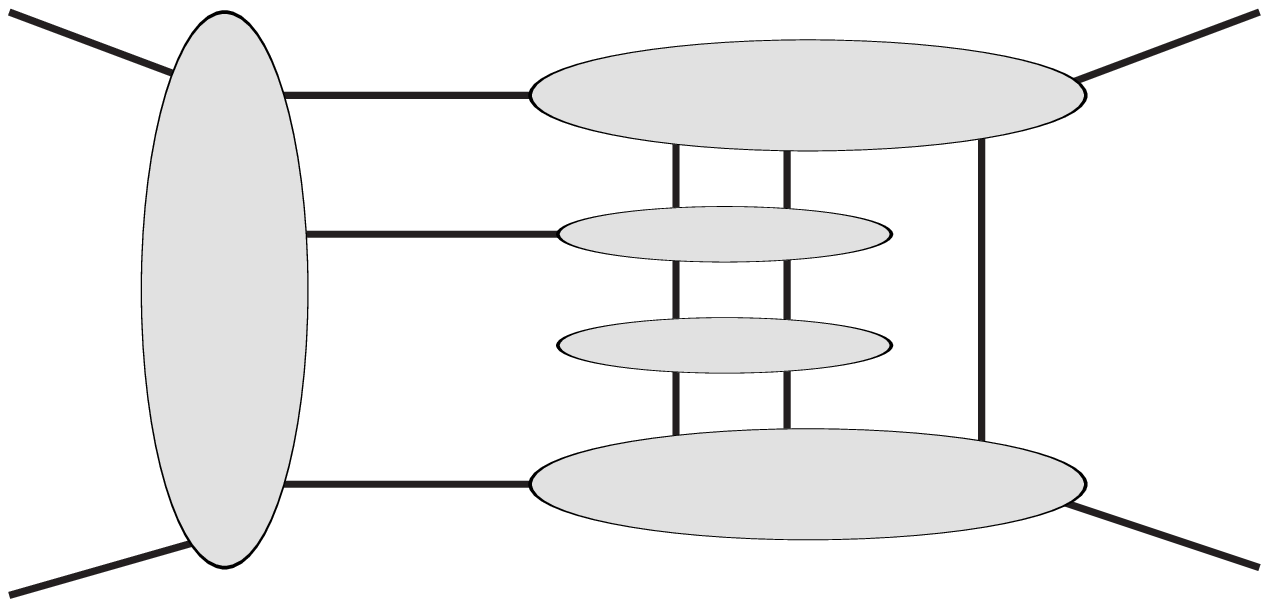}}
\caption{A nontrivial cut of the planar four-point six-loop
  $\NeqFour$ sYM theory.}
\label{SixLoopFigure}
\end{center}
\end{figure}

Another interesting case is the four-point six-loop amplitude of
$\NeqFour$ sYM theory.  At six loops, the planar amplitudes can be
expressed as a linear combination of integrals similar to the
four-loop form (\ref{FourLoopForm})~\cite{SixLoops}, except that the
numerator $N_k$ is a polynomial in the loop and external momenta of
degree ten instead of degree six.
In four dimensions, any Gram determinant $\det (p_i\cdot p_j)$
vanishes for $p_i$ and $p_j$ corresponding to any five independent
momenta, since there can be no more than four linearly independent
momenta.  For the six-loop four-point amplitude we
have a total of nine independent external and loop momenta.  Thus with
four-dimensional momenta and spinors in the cuts, the constructed
amplitude is trivially invariant under the shifts,
\begin{equation}
N_m \rightarrow N_m + a_m \det (p_i\cdot p_j) \,,
\end{equation}
although its form changes. The $a_m$
are constants.  If we impose dual conformal
symmetry~\cite{DualConformal,FourLoopCusp,FiveLoop,DualConfWI}, the
planar numerators are fixed with $a_m=0$.  What about higher
dimensions?  Iterated two-particle cuts are simple to evaluate in $D$
dimensions~\cite{BRY,BDDPR}, with the result that all $a_m$ detectable
in such cuts vanish.  As a more nontrivial check, we evaluated the
cut shown in \fig{SixLoopFigure} in six dimensions using the methods
described above and numerically compared it against the same cut
obtained via four-dimensional methods~\cite{FiveLoop,
SuperSum,FourLoop} and extended into six dimensions. 
Because this cut is composed of four- and five-point tree amplitudes,
its evaluation in six dimensions is similar to the four-loop cut
described in \app{FourLoopCutK}.
We find that  $a_m=0$ in six dimensions to match the cut \fig{SixLoopFigure}.
Although we did not check a spanning set of cuts, this
result strongly suggests that all $a_m$ vanish.

\section{Dual conformal properties in higher dimensions}
\label{DualConformalSection}

The above results suggest that integrands of $\NeqFour$ sYM
evaluated in six dimensions should
have simple dual conformal properties.
If this were to hold, it would lead to important restrictions on the
form of the integrands, guiding their construction in six dimensions,
or equivalently in four dimensions but with masses.

In four dimensions, the dual conformal boost
generators are~\cite{DualConfWI},
\begin{equation}
K^\mu=\sum^n_{i=1}
\left[2x_i^\mu x_i^\nu\frac{\partial}{\partial x_i^\nu}-x_i^2
 \frac{\partial}{\partial x_{i\mu}}\right] \,,
\label{FourDimDualConf}
\end{equation}
where $\mu=0,1,2,3$, and the dual coordinates $x$ are defined via,
\begin{equation}
x_{i,i+1} = x_i^\mu-x_{i+1}^\mu=p_i^\mu\,.
\end{equation}
The principal constraint on amplitudes
comes from the inversion symmetry $x_i^\mu
\rightarrow {x_{i\mu} / x_i^2}$. Since the integrals are all
dual translational invariant, they will also be dual conformal boost
invariant as long as they are invariant under inversion. This is
because the conformal boost is a combination of translations and
inversions.  Under the inversion, we have,
\begin{equation}
x_{ij}^2 \rightarrow \frac{x_{ij}^2}{x_i^2 x_j^2} \,, \hskip 1 cm
 {d^4 x_i} \rightarrow \frac{d^4 x_i}{x_i^{8}} \,. \hskip 1 cm
\end{equation}

We can extend this to six dimensions by adding extra terms 
to the generator, which in SO(1,3)$\times$SO(2) notation, is
\begin{equation}
\hat{K}^\mu=K^\mu 
+\sum^n_{i=1}\left[2x_i^\mu\left(n_i\frac{\partial}{\partial n_i}
+\tilde{n}_i\frac{\partial}{\partial \tilde{n}_i}\right)
+(n_i^2+\tilde{n}_i^2)\frac{\partial}{\partial x_{i\mu}}\right] \,,
\label{sixddual}
\end{equation}
where we explicitly include the sign from our choice of metric in the
extra-dimensional components.
 The extra-dimensional dual variables are
\begin{equation}
\nonumber n_i-n_{i+1}=p_i^4\,, \hskip 2 cm 
\tilde{n}_i-\tilde{n}_{i+1}=p_i^5 \,.
\end{equation}
This modification of the dual conformal generators is chosen 
so that the integrands maintain the same dual conformal weight 
as in four dimensions.  On the other hand, the integration
measure does depend on the dimension,
\begin{equation}
{d^4 x_i} \rightarrow \frac{d^4 x_i}{\hat x_i^{8}}\,,  \hskip 1 cm  
{d n_i} \rightarrow \frac{d n_i}{\hat x_i^{2}}\,, \hskip 1 cm  
{d \tilde n_i} \rightarrow  \frac{d \tilde n_i}{\hat x_i^{2}}\,,
\hskip 1 cm  
{d^6 \hat x_i} \rightarrow  \frac{d^6 \hat x_i}{\hat x_i^{12}}\,,
\end{equation}
where here the hats denote six-dimensional quantities, with $\hat{x}_i^2
=x_i^2 - n_i^2 -\tilde{n}_i^2$.  To maintain the invariance of the
amplitude, we therefore should truncate the integration to four
dimensions, which is equivalent to treating the extra-dimensional
components as fixed masses. Of course, even if we choose to integrate
over the extra dimensions, the fact that the integrands transform with
a definite dual conformal weight puts severe restrictions on their form.
 The SO(1,3)$\times$SO(2) decomposition is
especially convenient for treating the amplitudes as four-dimensional
but containing mass parameters, using the decomposition
(\ref{SixDimSpinors}) of six-dimensional spinors into four-dimensional
ones. This extension of dual conformal properties to higher dimensions is
thus related to the extension for the Higgs regulated
case~\cite{HiggsReg}.

Note that \eqn{sixddual} gives simply the first four components of the
six-dimensional dual conformal boost generator. The observed form
invariance of the four-point integrands (after dividing out the tree
amplitude) suggests that we can treat  the generator in a fully
covariant form, similar to \eqn{FourDimDualConf}, except we allow the
indices $\mu$ and $\nu$ to run over all six components.

Furthermore, we note that the dual conformal boost generator can also be
defined on the spinor variables. We start with the form of the dual conformal boost generator defined on the dual coordinates, and modify it to 
act also on spinors and to preserve the
constraint equations that define the dual coordinates,
\begin{eqnarray}
(x_i-x_{i+1})^{AB}&=&p_i^{AB}=\lambda_i^{Aa}\lambda_{ia}^{B}\,,\nn\\
 (x_i-x_{i+1})_{AB}&=&p_{iAB}=\tilde{\lambda}_{iA\dot{a}}\tilde{\lambda}_{iB}^{\dot{a}}\,.
\label{constraint}
\end{eqnarray}
This gives the generator in the full space $(\lambda,\tilde{\lambda}, x)$,
\begin{eqnarray}
\nonumber \hat K^\mu&=&\sum_{j=1}^n\biggl[2 \hat x^\mu_j \hat x^\nu_j\frac{\partial}{\partial \hat x_j^\nu}- \hat x_j^2\frac{\partial}{\partial \hat x_{j\mu}}+\frac{1}{2}\lambda^A_{ja}(\sigma^\mu)_{AB}(\hat x_j+\hat x_{j+1})^{BC}\frac{\partial}{\partial \lambda^C_{ja}} \nn\\
&& 
\hskip 2 cm 
\null  + \frac{1}{2}\tilde{\lambda}_{jA\dot{a}}(\tilde{\sigma}^\mu)^{AB}
(\hat x_j+ \hat x_{j+1})_{BC}
\frac{\partial}{\partial \tilde{\lambda}_{jC\dot{a}}}\biggr] \,.
\end{eqnarray}

As in $D=4$, we expect the dual conformal properties to be more 
transparent for
superamplitudes rather than ordinary amplitudes~\cite{DeltaIdentity}.
It is not difficult to check that the four-point tree superamplitude
transforms covariantly under a six-dimensional extension of dual
conformal transformations. Introducing the dual superspace
coordinates,
\begin{equation}
\theta^A_i-\theta^A_{i+1}=\lambda^{Aa}_i\eta_{ia}=q^A_i\,, \qquad\qquad
\tilde{\theta}_{iA}-\tilde{\theta}_{i+1A}=\tilde{\lambda}_{iA\dot{a}}\tilde{\eta}_{i}^{\dot{a}}=\tilde{q}_{iA}\,,
\end{equation} 
the four-point tree-level superamplitude in dual superspace is
\begin{equation}
\mathcal{A}_4^\tree= -i \frac{\delta^6(\hat{x}_{1}-\hat{x}_{5})
  \delta^4(\theta_1-\theta_{5})\delta^4(\tilde{\theta}_1-\tilde{\theta}_{5})}
{\hat{x}^2_{13}\hat{x}^2_{24}}.
\end{equation}
Using the usual dual conformal-inversion properties of the supercoordinates,
\begin{equation}
\theta^A_i\rightarrow\frac{\hat{x}_{iAB}}{\hat{x}_i^2}\theta^B_i\,,
\qquad \qquad 
\tilde{\theta}_{iA}\rightarrow\frac{\hat{x}^{AB}_{i}}{\hat{x}_i^2}\tilde{\theta}_{iB},
\end{equation}
the four-point amplitude transforms as,
\begin{equation}
\mathcal{A}_4^\tree\rightarrow (\hat{x}_1^2)^2 (\hat{x}^2_1\hat{x}_2^2\hat{x}_3^2\hat{x}_4^2)\mathcal{A}_4^\tree \,.
\end{equation}
The extra weight on $\hat{x}_1$ is due to the delta functions, which choose
a specific point to enforce the cyclic identification. 

It would of course be interesting to further study the implications of
dual conformal transformations on higher-dimensional integrands,
especially at higher points.  For example, one 
can show using BCFW recursion that under 
dual conformal inversions for $n\geq 4$, the six-dimensional
superamplitudes behave as~\cite{DualConformalProof},
\begin{equation}
\mathcal{A}_n^\tree \rightarrow (\hat{x}_1^2)^2 \Bigl(\prod^n_{i=1}
\hat{x}^2_i \Bigr)\mathcal{A}_n^\tree \,,
\end{equation}
where $\hat x_1$ is the special point taken to enforce momentum conservation
in the dual space.

\section{Comments and Conclusions}
\label{ConclusionSection}

In this paper, we developed six-dimensional
helicity~\cite{CheungOConnell} as a practical tool in unitarity-based
calculations of loop amplitudes.  As simple illustrations we described
the unitarity cuts of one-loop four-point amplitudes in QCD.  As more
sophisticated examples, we also described multiloop amplitudes in
$\NeqFour$ sYM theory, using also the DHS on-shell
superspace~\cite{DHS}.  To illustrate this, we worked out the
three-particle cut of a two-loop four-point amplitude in some detail,
before turning to higher-loop cases. Our results confirm that the
four-loop four-point $\NeqFour$ sYM and $\NeqEight$ supergravity
amplitudes obtained in refs.~\cite{GravityFour,FourLoop} using mainly
four-dimensional techniques are complete for $D\le 6$, as expected.

There are a number of obvious further applications and studies that
would be interesting to carry out.  Our six-dimensional approach can
be reinterpreted as a massive approach, by expressing six-dimensional
spinors and momenta in terms of four-dimensional quantities, with the
extra-dimensional pieces interpreted as mass terms.  For one-loop QCD
amplitudes, this should give forms of the cuts similar to the
massive approach for rational terms taken by Badger~\cite{Badger}. It
would be interesting to make the connection precise. More generally,
it would, of course, be interesting to apply the unitarity-based
six-dimensional helicity method described here to cases of
phenomenological interest in QCD and to compare them to other
approaches.

In general, the integrands of amplitudes can contain terms that vanish
when loop momenta are set to $D=4$, but are nonvanishing in higher
dimensions. This can make it dangerous to extrapolate results from
cuts in four dimensions to higher dimensions.  Similarly, in a massive
infrared regularization scheme~\cite{HiggsReg}, or when studying the
Coulomb branch of gauge theories, additional mass terms can be
present.  The six-dimensional helicity formalism is well suited for
systematically deriving such terms via unitarity.  Interesting cases
are two-loop $\NeqFour$ amplitudes at six and higher points where we
know such additional terms can appear~\cite{TwoLoopSixPoints}.  Even
at four points, the power counting in this theory is such that
starting at six loops contributions proportional to Gram determinants
that vanish in four dimensions but are non-vanishing in higher
dimensions in principle can appear, though our results indicate that
they do not.  The six-dimensional formalism may also shed light on the
peculiar pattern of vanishing and non-vanishing integral coefficients
in MHV and non-MHV amplitudes at higher points in the $\NeqFour$
theory~\cite{NMHVTwoLoop}, because six-dimensional helicity treats
these cases in a unified way.  In principle, it would be worthwhile to
evaluate cuts using even higher dimensions, but six-dimensional
methods push any potential lost terms to rather large numbers of loops
or legs.  For example, covariance and simple power counting
considerations suggest that four-point amplitudes in $\NeqFour$ sYM
theory will be free of terms that can vanish in six dimensions through
at least eight loops.

We noted that the expressions for the four-point planar integrands
through at least six loops do not appear to differ between four and
six dimensions.  This suggests that the integrands of amplitudes in
higher dimensions will also be constrained by extensions of dual
conformal symmetry.  Based on this, we proposed a six-dimensional
extension of the dual conformal symmetry generators.  We also noted
that the four-point tree amplitudes of $\NeqFour$ sYM theory have
simple properties under this extension, and proposed the general
property at $n$ points~\cite{DualConformalProof}.  Our six-dimensional
extension is connected to a recently proposed extension of the dual
conformal generators to massive amplitudes~\cite{HiggsReg}. It is also
likely connected to a ten-dimensional extension~\cite{OConnell}.  More
generally, it seems likely that extensions of the four-dimensional
structures identified in planar loop integrands~\cite{NimaBCFW} will
restrict the form in higher dimensions as well.  The BCJ relations
between planar and nonplanar contributions~\cite{BCJ,LoopBCJ} then
suggest that nonplanar integrands will also be restricted by
extensions of the four-dimensional structures.  This warrants further
study, especially for larger numbers of external legs.

In summary, six-dimensional helicity and its associated on-shell
superspace are additional useful tools both for carrying out loop
calculations in conjunction with the unitarity method and for
uncovering new structures in scattering amplitudes.

\acknowledgments

We thank Lance Dixon, Michael Douglas, Johannes Henn, Henrik
Johansson, Donal O'Connell, and especially Radu Roiban for many
stimulating discussions.  We thank Academic Technology Services at
UCLA for computer support.  This research was supported by the US
Department of Energy under contract DE--FG03--91ER40662.
J.~J.~M.~C. gratefully acknowledges the Stanford Institute for
Theoretical Physics for financial support.  HI's work is
supported by a grant from the US LHC Theory Initiative through NSF
contract PHY--0705682.

\appendix

\section{An Analytic Four-Loop Cut}
\label{FourLoopCutK}

In this appendix, we present a nontrivial cut of the four-loop
four-point amplitude of $\NeqFour$ sYM.  In particular, we
analytically evaluate the cut shown in \fig{4LoopCut}, which is a
nonplanar permutation of the cut (k) in \fig{CutBasisFigure}. Since
this cut consists of only four- and five-point subamplitudes, the
calculation is not significantly more difficult than the two-loop
example in \sect{TwoLoopFourPoint}.  
The cut is evaluated by carrying out the Grassmann integration, 
\begin{eqnarray}
 C^{\text{4-loop}} &=& \int \biggl( \prod_{i=1}^{7}d^2\eta_{l_i}d^2\tilde{\eta}_{l_i} \biggr) 
  \mathcal{A}_{5}^\tree(p_1,p_2,l_1,l_2,l_3)\,
  \mathcal{A}_{4}^\tree(-l_1,-l_2,l_4,l_5) \nonumber\\
  &&\hskip 2 cm \null
 \times\mathcal{A}_{4}^\tree(-l_4,-l_5,l_6,l_7)\,
  \mathcal{A}_{5}^\tree(p_3,p_4,-l_3,-l_6,-l_7) \,.
\end{eqnarray}
Here we have seven on-shell internal momenta labeled $l_i$, and four
component tree amplitudes. In this cut,
we can use the supermomentum delta functions
to localize six of the seven internal supercoordinates
$\eta_{l_i},\tilde{\eta}_{l_i}$. The remaining supercoordinate
integration is handled in the same way as in the example of
\sect{TwoLoopFourPoint}.

A solution to the supercoordinate delta-function constraints is
\begin{eqnarray}
 q_{l_1} &\rightarrow& -s^{-1}_{l_1l_2} \s l_1 \s l_2 q_{l_3}  \,,
  \qquad \tilde{q}_{l_1} \rightarrow -s^{-1}_{l_1l_1} \s l_1 \s l_2 \tilde{q}_{l_3}\,, \nonumber\\
 q_{l_2} &\rightarrow& -s^{-1}_{l_1l_2} \s l_2 \s l_1 q_{l_3} \,,
  \qquad \tilde{q}_{l_2} \rightarrow -s^{-1}_{l_1l_2} \s l_2 \s l_1 \tilde{q}_{l_3}\,, \nonumber\\
 q_{l_4} &\rightarrow& -s^{-1}_{l_4l_5} \s l_4 \s l_5 q_{l_3} \,,
  \qquad \tilde{q}_{l_4} \rightarrow -s^{-1}_{l_4l_5} \s l_4 \s l_5 \tilde{q}_{l_3} \,,\nonumber\\
 q_{l_5} &\rightarrow& -s^{-1}_{l_4l_5} \s l_5 \s l_4 q_{l_3} \,,
  \qquad \tilde{q}_{l_5} \rightarrow -s^{-1}_{l_4l_5} \s l_5 \s l_4 \tilde{q}_{l_3} \,,\nonumber\\
 q_{l_6} &\rightarrow& -s^{-1}_{l_6l_7} \s l_6 \s l_7 q_{l_3} \,,
  \qquad \tilde{q}_{l_6} \rightarrow -s^{-1}_{l_6l_7} \s l_6 \s l_7 \tilde{q}_{l_3} \,,\nonumber\\
 q_{l_7} &\rightarrow& -s^{-1}_{l_6l_7} \s l_7 \s l_6 q_{l_3} \,,
  \qquad \tilde{q}_{l_7} \rightarrow -s^{-1}_{l_6l_7} \s l_7 \s l_6 \tilde{q}_{l_3}\,,
\end{eqnarray}
where we have ignored all dependence on $\{q_1,q_2,q_3,q_4\}$, since
these will drop out after the final $\eta_{l_3}$ integration. As usual,
we take the Mandelstam invariants to be $s_{l_il_j} =(l_i + l_j)^2$. Because
of extra factors of these invariants coming from the
Grassmann integrations --- seen in
\eqn{local} --- we must also multiply the final cut by
$s^2_{l_1l_2}s^2_{l_4l_5}s^2_{l_6l_7}$. The rest of the calculation
is similar to the derivation of \eqn{TwoLoopCutResult}, giving,
\begin{eqnarray}
 C^{\text{4-loop}} &=& \frac{ s_{23}(l_4+l_5)^2 (l_6+l_7)^2\mathcal{A}_{4}^{\tree}(p_1,p_2,p_3,p_4)}
    {s_{12} (p_1+l_3)^2 (p_2+l_1)^2 (p_3-l_7)^2 (p_4-l_3)^2 (l_2+l_3)^2 (l_2-l_4)^2 (l_5-l_6)^2 (l_3+l_6)^2} \nonumber\\
  &&\null \times\langle l_3^a|   \biggl( \s p_1\s p_2\s l_1\s l_2 + \frac{\s l_2\s l_3\s p_1\s p_2\s l_1\s l_2+\s l_1\s l_2\s l_3\s p_1\s p_2\s l_1}{(l_1+l_2)^2} \nonumber\\
    &&\qquad\null - \frac{\s l_1\s l_2(\s l_3\s p_1\s p_2\s l_1-\s l_3\s l_1\s p_2\s p_1) - (\s l_1\s p_2\s p_1\s l_3-\s p_1\s p_2\s l_1\s l_3)\s l_2\s l_1}{2(l_1+l_2)^2}\biggr)   |l_{3\da}] \nonumber\\
  &&\null \times\langle l_{3a}| \biggl( \s l_6\s l_7\s p_3\s p_4 + \frac{\s l_7\s p_3\s p_4\s l_3\s l_6\s l_7 + \s l_6\s l_7\s p_3\s p_4\s l_3\s l_6}{(l_6+l_7)^2} \nonumber\\
    &&\qquad \null+\frac{(\s l_7\s p_3\s p_4\s l_3-\s l_7\s l_3\s p_4\s p_3)\s l_7\s l_6 - \s l_6\s l_7(\s l_3\s p_4\s p_3\s l_7-\s p_3\s p_4\s l_3\s l_7)}{2(l_6+l_7)^2}  \biggr) |l_3^{\da}] \,.
\label{4LoopAnalytic}
\end{eqnarray}
This expression is just a gamma trace, and although it appears to be
chiral, the $\gamma_7$ term in the trace actually drops out. In fact,
the five-point tree amplitude itself is non-chiral, although this
property is not manifest in the present form.

One can compare this to the cut of the amplitude derived in 
ref.~\cite{FourLoop} using (mostly) four-dimensional methods,
\begin{eqnarray}
 C^{\text{4-loop}} &=& \frac{s_{12}s_{23}\mathcal{A}_4^{\tree}(p_1,p_2,p_3,p_4)}{(p_2+l_1)^2(p_3-l_7)^2(l_1-l_5)^2(l_4-l_7)^2} \nonumber\\ 
    &&\times\Biggl(\frac{s_{12}^2(l_2-l_6)^2}{(l_2+l_3)^2(l_3+l_6)^2} + \frac{s_{23}(l_4+l_5)^4}{(p_1+l_3)^2(p_4-l_3)^2} \nonumber\\
    &&\
\null \hskip .5 cm 
 + \frac{s_{12}(l_4+l_5)^2(p_1+l_3+l_6)^2}{(p_1+l_3)^2(l_3+l_6)^2} + \frac{s_{12}(l_4+l_5)^2(p_4-l_2-l_3)^2}{(p_4-l_3)^2(l_2+l_3)^2}
\Biggr) \,.
\end{eqnarray}
We have evaluated these two expressions numerically using
six-dimensional momenta to verify their equality.

\section{Auxiliary variables at three points}
\label{ThreePointAppendix}

In this appendix, we summarize the construction of the variables needed
to define the three-point amplitudes in
\eqns{ThreePointAmplitude}{supertrees}, as given in
ref.~\cite{CheungOConnell}.  At three points, momentum conservation and
on-shell conditions mean that the Lorentz products of the momenta all
vanish.  This gives a vanishing determinant of the Lorentz-invariant
inner product of a fundamental and an anti-fundamental spinor,
i.e. the 2$\times$2 matrix $\langle i_a|j_{\dot{a}}]$ has rank one, so
that $\langle i_a|j_{\da}]=u_{ia}\tilde{u}_{j\dot{a}}$.  Consistently
defining these SU(2) spinors for all Lorentz invariants, we have (for
$\{i,j\}$ cyclically ordered),
\begin{eqnarray}
\langle i_a|j_{\dot{a}}]=u_{ia}\tilde{u}_{j\dot{a}}\,, \hskip 1 cm && \hskip 1 cm 
    \langle j_a|i_{\dot{a}}]=-u_{ja}\tilde{u}_{i\dot{a}}\,.
\label{not}
\end{eqnarray}
One important property of these SU(2) spinors can be derived from
momentum conservation, 
\begin{eqnarray}
u_1^{c}\langle1_{c}|=u_2^{c}\langle2_{c}|=u_3^{c}\langle3_{c}|\,,
\hskip 1 cm && \hskip 1 cm \tilde{u}_{1\dot{c}}[1^{\dot{c}}|=\tilde{u}_{2\dot{c}}[2^{\dot{c}}|=\tilde{u}_{3\dot{c}}[3^{\dot{c}}|\,.
\label{law1}
\end{eqnarray}
Ref.~\cite{CheungOConnell} also defines a pseudoinverse of these
spinors,
\begin{eqnarray}
u_{a}w_{b}-u_{b}w_{a}=\epsilon_{ab} \,,\hskip 1 cm && \hskip 1 cm \tilde{u}_{\dot{a}}\tilde{w}_{\dot{b}}-\tilde{u}_{\dot{b}}\tilde{w}_{\dot{a}}=\epsilon_{\dot{a}\dot{b}}\,,
\label{learn}
\end{eqnarray} 
subject to the additional condition,
\begin{eqnarray}
\hskip 0.5 cm w_1^a\lambda^A_{1a}+w_2^a\lambda^A_{2a}+w_3^a\lambda^A_{3a}=0\,,
\hskip 1 cm && \hskip 1 cm \tilde{w}_{1\dot{a}}\tilde{\lambda}_{1A}^{\dot{a}}+\tilde{w}_{2\dot{a}}\tilde{\lambda}_{2A}^{\dot{a}}+\tilde{w}_{3\dot{a}}\tilde{\lambda}_{3A}^{\dot{a}}=0 \,. 
\label{law2}
\end{eqnarray}


\end{document}